\documentclass[11pt]{article}
\usepackage{fullpage}
\usepackage{amsmath}
\usepackage{amssymb}
\usepackage{amsthm}
\usepackage{graphicx}
\usepackage[english]{babel}
\usepackage[babel=true]{csquotes}
\usepackage{wrapfig}
\usepackage{float}
\usepackage{epstopdf}

\usepackage{graphics,subfig}
\usepackage{epsfig}

\usepackage{booktabs}
\usepackage{enumerate}
\usepackage{epsfig}
\usepackage{pxfonts}

\usepackage{color}
\usepackage[numbers]{natbib}
\usepackage{hyperref}
\usepackage{tikz}
\usepackage{appendix}
\usepackage{soul}
\usepackage{multirow}
\usepackage{graphicx}
\usepackage{caption}
\usepackage{algpseudocode}
\usepackage{algorithm}

\usepackage{amsmath, amsthm, amssymb}

\usepackage{amsthm}
\graphicspath{{./figures/}}

\usetikzlibrary{%
decorations.fractals,%
decorations.shapes,%
decorations.text,%
decorations.pathmorphing,%
decorations.pathreplacing,%
decorations.footprints,%
decorations.markings}

\newcommand{\ra}[1]{\renewcommand{\arraystretch}{#1}}

\newtheorem{theorem}{Theorem}[section]
\newtheorem{lemma}[theorem]{Lemma}
\newtheorem{fact}[theorem]{Observation}
\newtheorem{proposition}[theorem]{Proposition}
\newtheorem{prop}[theorem]{Proposition}
\newtheorem{corollary}[theorem]{Corollary}

\newtheorem{remark}[theorem]{Remark}

\usepackage{soul}
\makeatletter

\newcommand{\Rmnum}[1]{\expandafter\@slowromancap\romannumeral #1@}
\def\BState{\State\hskip-\ALG@thistlm}
\makeatother

\newcommand{\vh}[1]{{\textcolor{black}{ #1}}}
\begin{document}
\title{Container Relocation Problem: \\Approximation, Asymptotic, and Incomplete Information
}
\author{Setareh Borjian
       \thanks{ Oracle, Burlington.
    Email: \protect \url{setareh.borjian.boroujeni@oracle.com}.}
\and
Virgile Galle
    \thanks{ ORC, MIT.
    Email: \protect\url{vgalle@mit.edu}.}
\and
Vahideh H. Manshadi
    \thanks{ School of Management, Yale.
    Email: \protect\url{vahideh.manshadi@yale.edu}.}
\and
Cynthia Barnhart
    \thanks{ CEE, MIT.
    Email: \protect\url{cbarnhar@mit.edu}.}
\and
Patrick Jaillet
    \thanks{ EECS and ORC, MIT.
    Email: \protect\url{jaillet@mit.edu}.}
}
\date{}
\maketitle

\begin{abstract}
The Container Relocation Problem (CRP) is concerned with finding a sequence of moves of containers that minimizes the number of relocations needed to retrieve all containers respecting a given order of retrieval. While the problem is known to be NP-hard, there is much evidence that certain algorithms (such as the $A^*$ search \cite{Astar3}) and heuristics
perform reasonably well on many instances of the problem.

In this paper, we first focus on the $A^*$ search algorithm, and analyze lower and upper bounds that are easy to compute and can be used to prune nodes and \vh{also} to determine the gap between the solution found and the optimum. Our analysis sheds light on which bounds result in fast computation within a given approximation gap. We also present extensive simulation results that improve upon our theoretical analysis, and further show that our method finds the optimum solution on most instances of medium-size bays. On the ``hard'' instances, our method finds an approximate solution with a small gap and within a time frame that is fast for practical applications. We also study the average-case asymptotic behavior of the CRP where the number of columns grows. We calculate the expected number of relocations in the limit, and show that the optimum number of relocations converges to a simple and intuitive lower-bound. This gives strong evidence that the CRP is ``easier'' in large instances, and heuristics such as \cite{RePEc} finds near optimal solution.

We further study the CRP with incomplete information by relaxing the assumption that the order of retrieval of all containers are initially known. This assumption is particularly unrealistic in ports without an appointment system. In our model, we assume that the retrieval order of a subset of containers is known initially and the retrieval order of the remaining containers is observed later at a given specific time. Before this time, we assume a probabilistic distribution on the retrieval order of unknown containers. We extend the $A^*$ algorithm and combine it with sampling technique to solve this two-stage stochastic optimization problem. We show that our algorithm is fast and the error due to sampling and pruning the nodes is reasonably small. Using our framework, we study the value of information and its timing effect on the average number of relocations.

\end{abstract}

\section{Introduction}
With the growth in international container shipping in maritime ports, there has been an increasing interest in improving the operations efficiency in container terminals on the sea side and land side. The operations on the sea side include loading export containers on the vessels or discharging import containers from vessels and loading them onto internal trucks. The import containers are then transferred to the land side and are stacked in the storage area. These containers are later retrieved and delivered to external trucks to be distributed in the city.

\begin{figure}[h]
\centering
\includegraphics[width=0.25\textwidth]{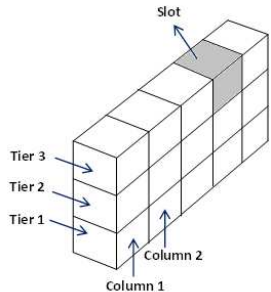}
\caption{Illustration of a bay of containers in storage area}
\label{fig:Bay}
\end{figure}

Due to limited space in the storage area, containers are stacked in tiers on top of each other. As shown in Figure \ref{fig:Bay}, several columns of containers in a row create a bay of containers. If a container that needs to be retrieved (\textit{target} container) is not located at a top most tier and is covered by other containers, the blocking containers must be relocated to another slot. As a result, during the retrieval process, one or more relocation moves are performed by the yard cranes. Such relocations are costly for the port operators and result in delay in the retrieval process. Thus, reducing the number of relocations is one of the main challenges faced by port operators. Finding the sequence of moves that minimizes the number of relocations while retrieving containers from a bay in a pre-defined order is referred to as the Container Relocation Problem (CRP) or the Block Relocation Problem (BRP).
A common assumption of the CRP is that only the containers that
are blocking the target container can be relocated. We refer to the CRP with this setting
as the \textit{restricted CRP}.

The CRP has been widely studied in the literature and has been shown to be NP-hard \cite{RePEc}. A few papers have developed mathematical models for the CRP and have solved small-sized instances of the problem exactly (\cite{RePEc}, \cite{ref9}, \cite{ref7}). In other papers, different heuristics are designed and tested on real-sized instances (\cite{ref5}, \cite{ref8}, \cite{ref4}, \cite{ref12}). Many of these heuristics are fast and shown to be empirically close to optimal for small-sized instances. However, none of them have provable guarantees and their performance on large-sized instances cannot be well assessed. Among the optimization methods that provide exact solution for the CRP, the $A^*$ algorithm (a branch and bound framework) has shown to be very effective and able to solve large instances of the problem in a time frame that is suitable for practical application \cite{Astar3}, \cite{Astar1}, \cite{Astar2}.

In this paper, we apply the $A^*$ algorithm (\cite{Astar3}, \cite{Astar1},\cite{Astar2}) to the CRP, and analyze various upper and lower bounds. First, we study the properties of a \vh{counting} lower bound  \vh{that simply counts the number of blocking containers in the initial configuration} (\vh{introduced in }\cite{ref5}). We show that this bound is non-decreasing on each path of the $A^*$ tree, implying that the guaranteed optimality gap given by the $A^*$ algorithm may only be improved as more levels of the tree are constructed (see Proposition \ref{prop:path1}).
Moreover, we \vh{analyze} \vh{a generalized} family of \vh{ {\em look-ahead lower bounds}} \vh{that also counts the future relocation moves due to unavoidable blocking}  (\vh{introduced in } \cite{Astar1}), \vh{and we prove that this family provide tighter lower-bounds that are still monotone in each path (see Proposition \ref{prop:LB decreasing} and \ref{prop:augmentingpathLN} and Figure \ref{fig:Exp3}). Our result implies that using look-ahead lower bounds (with a slightly  higher computation time compared to the counting bound) can decrease the number of branching needed to achieve a desired optimality gap.}

We also study several properties of the H heuristic presented in \cite{RePEc} \vh{which indicates that this heuristic serves well as an upper bound}.
In particular, we prove that the $A^*$ algorithm terminates after a certain level when the H heuristic is used
\vh{combined with the appropriate lower bound} (see Proposition \ref{prop:caseC1containers}, \ref{prop:caseCcontainers}, and \ref{prop:caseCkcontainers}).
\vh{Our numerical simulations show that, }
the algorithm typically finds the optimal solution much faster than what is suggested by the theoretical results (see Figures \ref{fig:Exp1} and \ref{fig:Exp2}).
We also use the idea of branching to design a heuristic called the Tree-Heuristic (TH-L, see Algorithm \ref{algo2}). We use randomly generated instances to benchmark several existing heuristics and the \vh{new} Tree Heuristic against the optimal solution given by the $A^*$ algorithm. \vh{We} show that TH-L has the best performance (smallest optimality gap) on average (see Table \ref{ref:bench}).

Building upon the previous results, we perform an average case analysis of the CRP \vh{in a bay with finite height.}
First we compute the expected \vh{value of the counting} lower bound (\cite{RePEc}, see Proposition \ref{lemmalinlowerbound}) \vh{for an evenly distributed bay}.
Then we provide an asymptotic result and \vh{prove that as the number of columns grows, in expectation, the optimum number of relocations converges to the counting lower bound}
(see Theorem \ref{theor:asym}). The intuition behind this result is that as the number of columns grows, the probability of finding a “good” column for a blocking container (i.e., a column where the blocking container will never be relocated again) approaches 1. In that case, the optimal number of relocations is essentially the number of blocking containers in the initial bay. Furthermore, we strengthen this result by showing experimentally that the expected number of relocations performed by heuristic H (\cite{RePEc}) \vh{also converges to the counting lower bound}
(see Figure \ref{fig:Exp4a}). This \vh{further} highlights the effectiveness of both heuristics H and TH-L in the case of large instances.

A critical assumption of the CRP is that the retrieval order of all containers in the bay is considered to be known in advance. However, in practice, such information is not available far in advance and the exact departure order of containers becomes known only as their retrieval times approach.
\vh{Here we relax this assumption in the following way: We consider} a 2-stage setting, where the retrieval order of a subset of containers is known initially and the retrieval order of the remaining containers is revealed all at once at a later time. \vh{We refer to this problem as the {\em CRP with incomplete information}.}
Before the entire information becomes available, we assume a probabilistic distribution on the \vh{unknown} retrieval orders.
We develop a 2-stage approximate stochastic optimization algorithm, called ASA*, and use it to solve medium-sized instances of the CRP with incomplete information. This algorithm is an extension of the $A^*$ algorithm combined with sampling and pruning techniques \vh{based on the} bounds defined \vh{earlier}.
We give theoretical bounds on the approximation error \vh{incurred by}
this algorithm (see Propositions \ref{pro1}, \ref{pro2}, and \ref{pro3}). Furthermore, we introduce a myopic heuristic, which is fast and performs well on average compared to the ASA* algorithm (see Figure \ref{gaps}). We use the heuristic to solve large instances of the problem and show through experiments that \vh{the ratio of expected number of relocations in the incomplete information setting to that of complete information}
converges to a constant, as the bay gets larger (see Figure \ref{infoBays}). Moreover, we use the ASA* algorithm and the myopic heuristic to assess the value of information (see Figure \ref{gaps}). Our experiments show that when the retrieval order of 50\%-75\% of containers is known at the beginning, the loss from the missing information is negligible and the average number of relocations is very close to that of the CRP with full information.

\vh{Our algorithms can serve as a decision support tool for port operators as well as an analytical tool to draw insights on managerial decisions such as determining the size/capacity of the bays and assessing the value of information. }
This issue of lack of information is \vh{particularly} critical as it captures port operations more realistically. Understanding the value of information could help port operators design efficient appointment systems for trucks.

\subsection{Literature Review}

\vh{Stahlbock  and Vo\ss \cite{ref17} provide a general survey on port operations literature.}
For a recent general review and classification survey of the existing literature on the CRP as well as the stacking problem, \vh{we refer the reader to} Lehnfeld and Knust \cite{Survey2014}.

Many heuristics have been developed for the CRP. Kim and Hong suggest
a decision rule that uses an estimate of expected number of additional relocations for a column \cite{ref5}.
Caserta et al. \cite{ref8} propose a binary description of configurations and use it in a four steps
heuristic. Lee and Lee \cite{ref4} consider an extended objective incorporating working time of cranes in a
three phases heuristic. Caserta et al. \cite{ref12} derive a meta-heuristic based on the
corridor method, which again optimizes over a restricted set of moves and Caserta et al. in
\cite{RePEc} present a heuristic that is defined later in this paper. \"{U}nl\"{u}yurt and Ayd{\i}n \cite{ref22} propose a branch and bound approach with several heuristics based on this idea. Foster and Bortfeld \cite{ref23} use a tree search procedure to improve a greedy initial solution.
Finally, Petering and Hussein \cite{ref9} describe a look-ahead heuristic that
does not make the common assumption of the restricted CRP. \vh{In our experimental result, we use randomly generated instances
to compare the performance of these heuristics (and our new TH-L heuristic) to that of the optimal
solution given by the A* algorithm. We show that on average our TH-L outperforms the existing heuristics. }

In the mathematical programming formulations that
insure optimality on the entire feasible set of moves, the first binary formulation was suggested
by Kim and Hong \cite{ref5}. Petering and Hussein \cite{ref9} propose another formulation which is more
tractable but cannot solve real-sized instances efficiently. Caserta et
al. \cite{RePEc} derive a different Integer Program (IP). Borjian et al. \cite{sb} develop an IP formulation which allows for incorporating a time component into the model. This model is used to jointly optimize the number of relocations and total service delay.
In all these IP formulations, due to the combinatorial nature of the problem, the number of variables and constraints dramatically increases as the size of the bay grows, and the IP cannot be solved for large instances. A way to bypass this problem has been to
look at a new type of algorithm called $A^*$. It was first introduced to this
problem by Zhang et al. \cite{Astar3}, studied by Zhu et al. \cite{Astar1}, and Tanaka and Takii \cite{Astar2} proposed
a new lower bound for the algorithm. \vh{Our paper contributes to this line of research (applying $A^*$ to CRP) by theoretically analyzing several lower and upper bounds, also by computing (both theoretically and via simulation) how the optimality gap behaves as the number of branches increases.}

\vh{Average case analysis of CRP is fairly new. The only other paper in this direction is the recent paper by Olsen and Gross \cite{avgcaseanalysis}. They }
also provide \vh{a probabilistic}
analysis of the asymptotic CRP when \vh{both}
the number of columns \vh{and tiers} grow to infinity.
They show that there exists a polynomial time algorithm that solves this problem close to optimality with high probability.
\vh{Our model departs from theirs in two (related) aspects: (i) We keep the maximum height (number of tiers) a constant whereas in \cite{avgcaseanalysis} the height also grows. Our assumption is motivated by the fact that the maximum height is limited by the crane height, and it cannot grow arbitrarily. (ii) We assume the ratio of the number of containers initially in the bay to the bay size stays constant (i.e., the bay is almost full at the beginning). On the other hand, in \cite{avgcaseanalysis}, the ratio of the number of containers initially in the bay to the bay size decreases (and it approaches zero) as the number of columns grows. In other words, in the model of \cite{avgcaseanalysis}, in large bays, the bay is under-utilized.}

Finally, we mention that another major challenge in port optimization is uncertainty (together with the value of
information) which has not been considered in the literature until very recently. Zhao and Goodchild
\cite{RePEc:eee:transe:v:46:y:2010:i:3:p:327-343}
model uncertainty in truck arrivals at the terminal. They assume
that trucks arrive in groups and that the information is revealed for the
whole group at the same time. Using this assumption of rolling information, they
use
\vh{a myopic heuristic }
to get insights on the trade-off between information and
optimality.
Ku \cite{dusan} consider a model where the trucks arrive in groups, and the retrieval order within in each group is unknown in advance. 
They propose a stochastic dynamic programming for this problem. They note that the size of decision tree grows exponentially with the bay size, making the computation time prohibitive. 
Thus they develop a heuristic called Expected Reshuffling Index (ERI) to approximately solve the problem. This heuristic, moves each blocking container to the column that minimizes the expected number of future relocations.
\vh{In our work, we model uncertainty as a 2-stage stochastic optimization problem, and we use a generalization of the $A^*$ algorithm combined with sampling and pruning techniques to find the near optimal solution of this stochastic optimization problem.}


This paper is structured as follows: Section 2 presents the $A^*$
Algorithm in the full information case. It describes the method,
\vh{introduces}
lower and upper bounds and derives theoretical results for these bounds.
Section 3 presents the asymptotic analysis of the C\@R\@P.
Section 4 provides a stochastic scheme for the case of incomplete information
and studies the value of information about container departure times.
In Section 5, we present several experimental results
\vh{for the}
average case
analysis of the C\@R\@P in both the complete and incomplete information cases.
Finally, Section 6 concludes and suggests future research topics.
The proofs for all
\vh{theoretical statements}
are given in the Appendix.

\section{Optimal solution for the CRP with full information}
\label{sec:SDT}
In this section and for the sake of completeness, we present the $A^*$ algorithm applied to
the CRP, as previously introduced by~\cite{Astar1}. We also discuss bounds developed in the literature
(\cite{RePEc} and~\cite{Astar1}) and derive new theoretical results on them.

\subsection{Description of the algorithm}
\label{descalgo}
Before describing the $A^*$ algorithm, we review the Container Relocation Problem (CRP) and introduce a few notations used throughout the paper:  We are given a bay $B$ with $C$ columns and $P$ tiers, where $C \geq P \geq 3$. Initially $N$ containers are stored in the bay. We label the containers based on their departure order, i.e., container 1 is the first one to be retrieved. The CRP corresponds to finding a sequence of moves to retrieve containers $1, 2, \ldots, N$ (respecting the order) with a minimum number of relocations.  For bay $B$, we denote the minimum number of relocation by $z_{opt}(B)$.

To find a solution, we use the $A^*$ algorithm, which is basically a move-based approach
using a special decision tree. This approach takes 4 entries: $B$ is the bay with $N$ containers to be solved optimally.
$R(D)$ and $S(D)$ are functions giving
respectively upper and lower bounds on $z_{opt}(D)$ for any configuration D. In this paper, upper bound
functions that are considered also give us feasible solutions. Finally, $\mathcal{N}$ is the
maximum number of nodes in the decision tree.

The algorithm returns two results: The best incumbent (in objective function) found by $A^*$, denoted $z_A$, to which we associate the sequence of moves needed to find this incumbent called $\sigma_A$ and the gap
guaranteed between $z_A$ and
$z_{opt}(B)$ denoted by $\gamma$. Notice that if the algorithm did not reach $\mathcal{N}$ nodes, then the algorithm found an solution (i.e. $\gamma=0$ and $z_A=z_{opt}(B)$).

Let us define the following notations: $l$ is the level of the tree. A node is at level $l$, if
$l$ relocations have been done from the initial bay to get to this node. This is the reason why this method is
called a move-based approach. Below we describe the algorithm in detail and introduce the notation used in the description of Algorithm~\ref{algo}.

\begin{algorithm}[htb!]
\caption{$A^*$ Algorithm}\label{algo}
\begin{algorithmic}[1]
\Procedure{[$z_A$,$\gamma$]=$A^*(B,R,S,\mathcal{N})$}{}
\BState\ \emph{Pre-processing}:
\While{target container $n$ is on top}
  \State{retrieve target container $n$ of $B$ and } $n \gets n+1$
\EndWhile\
\BState\ \emph{Initialize}:
\State\ $z_A \gets \infty$, $l \gets 0$, $B^l \gets B$, $m \gets 0$ (number of nodes)
\While{there exists a non-visited node}
  \For{all nodes $B^l$ at level $l$}
  \State\ Mark node $B^l$ as visited
  \State\ $U(B^l)=R(B^l)+l$ and $L(B^l)= S(B^l)+l$
\BState\ \emph{Updating the incumbent:}
    \If{$U(B^l)<z_A$}
      \State\ $z_A \gets U(B^l)$
   \EndIf\
\BState\ \emph{Pruning:}
   \If{$U(B^l) \le L(B^l)$ or $L(B^l) \ge z_A$}
       \State\ Prune node $B^l$ (i.e. stop branching from this node)
    \Else
\BState\ \emph{Branching:}
      \For{Every ``Child'' of $B^l$}
        \If{$m \ge \mathcal{N}$}
          \State\ \textbf{Stop}
        \Else\
          \State\ add a non-visited ``Child'' to the tree at level $l+1$ and $m=m+1$
        \EndIf\
      \EndFor\
    \EndIf\
  \EndFor\
  \State\ $l \gets l+1$
\EndWhile\
\BState\ \emph{Gap:}
\State\ $L_{min} = \underset{\textit{D non-visited leaf}} {\min}(L(D))$
\State\ $\gamma \gets z_A - L_{min}$
\EndProcedure\
\end{algorithmic}
\end{algorithm}

The tree represents the sequence of relocation moves. Suppose $n$ is the target container. If $n$ is not blocked, it
can be retrieved since a retrieval is optimal. In that case, no decision has to be made, therefore no new node
is created. If $n$ is blocked, the topmost blocking container needs to be
relocated to another column. This column has to be chosen in order to minimize all future
relocations. In order to do so, we generate all possible feasible moves for the blocking container by
creating children of the corresponding node in a decision tree procedure. Considering all possible nodes gives
the optimal solution.

 Building the whole tree of feasible moves would not be tractable since the number
 of nodes increases exponentially with the number of relocations. Consequently,
 we use certain decision rules to prune the tree with the guarantee of
 keeping the optimal path in the tree.

Let $B^l$ be any configuration that appears at level $l$ in the
tree. By definition,
\begin{eqnarray}
S(B^l) \le z_{opt}(B^l) \le R(B^l)
\end{eqnarray}
Let us define the cumulative upper and lower bounds as the sum of the bound evaluated at the current state and the number of relocations already performed (i.e., $l$ relocations have been done at level $l$ of the tree) $U(B^l)=R(B^l)+l$ and $L(B^l)= S(B^l)+l$ (line 11 of
Algorithm~\ref{algo}).
Notice that in this case
\begin{eqnarray}
L(B^l) \le z_{opt}(B) \le U(B^l) \textit{, for all } B^l \textit{ in the $A^*$ tree}
\end{eqnarray}

At any level $l$, the algorithm prunes paths using the following two rules (line 17 of
Algorithm~\ref{algo}):
 \begin{enumerate}[(i)]
  \item If $L(B^l)= U(B^l)$, then $z_{opt}(B^l)=U(B^l)$ and we can stop
  branching on this path. From
  this point, we can simply follow the feasible solution given by the upper
  bound.
\item If $L(B^l) \ge z_A$ (i.e. the lower bound $L(B^l)$ is larger than the best solution found so far $z_A$) then this node can be pruned because the
optimal solution for this node is going to be greater than its lower bound, hence greater than $z_A$.
The best incumbent is updated if a $B^l$ such that $U (B^l)<z_A$ is found (lines 13, 14 of
Algorithm~\ref{algo}).
\end{enumerate}
By construction, the above rules maintain the
optimal path in the $A^*$ tree.

\begin{figure}[htdp]
\begin{center}
\begin{tabular}{|c|c|c|} \hline
 6 &   &   \\ \hline
 1 & 5 &   \\ \hline
 4 & 2 & 3 \\ \hline
\end{tabular}
\end{center}
\caption{Bay with 3 tiers, 3 columns and 6 containers}
\label{tab:ex}
\end{figure}

\tikzstyle{level 1}=[level distance=5cm, sibling distance=5cm]
\tikzstyle{level 2}=[level distance=3.5cm, sibling distance=2cm]

\tikzstyle{bag} = [text width=5em, text centered]
\tikzstyle{end} = [circle, minimum width=3pt,fill, inner sep=0pt]

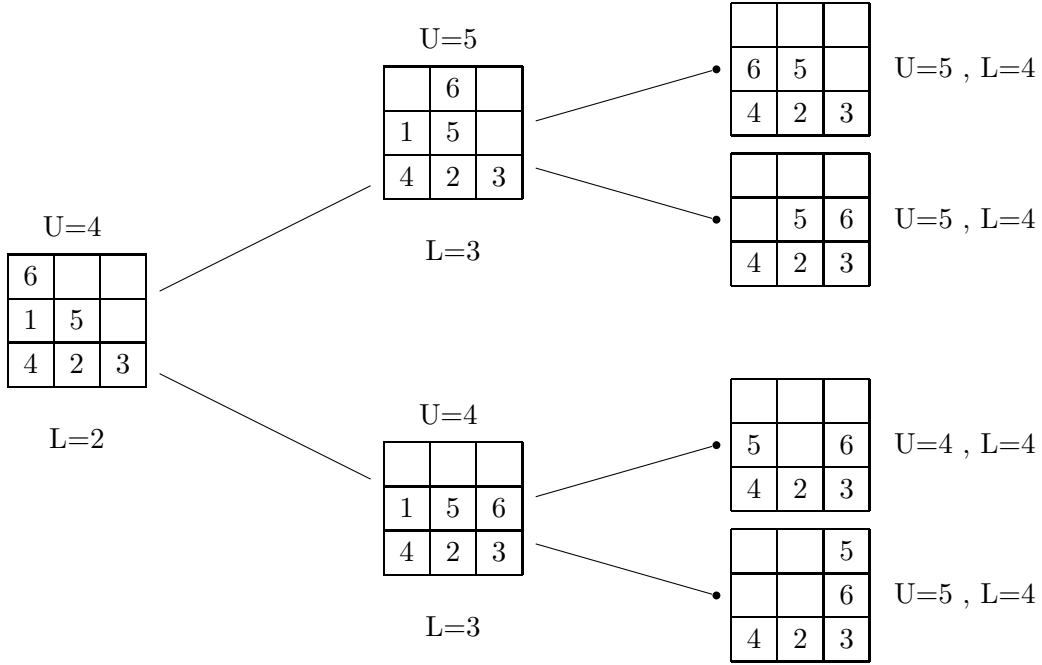
\begin{figure}
\centering
\begin{tikzpicture}[grow=right, sloped][Example of Smart Decision Tree]
\node[bag] {U=4
\vspace{2mm}
\\
\begin{tabular}{|c|c|c|} \hline
 6 &   &   \\ \hline
 1 & 5 &   \\ \hline
 4 & 2 & 3 \\ \hline
\end{tabular}
\vspace{2mm}
\\
L=2}
    child {
        node[bag] {
        U=4
        \vspace{2mm}
\\
\begin{tabular}{|c|c|c|} \hline
  &   &   \\ \hline
 1 & 5 &  6 \\ \hline
 4 & 2 & 3 \\ \hline
\end{tabular}
\vspace{2mm}
\\
L=3}
            child {
                node[end, label=right:
                    {\begin{tabular}{|c|c|c|} \hline
  &   &  5 \\ \hline
  &  &  6 \\ \hline
 4 & 2 & 3 \\ \hline
\end{tabular}\hspace{2mm} U=5 , L=4}] {}
                edge from parent
                node[above] {}
                node[below]  {}
            }
            child {
                node[end, label=right:
                    {\begin{tabular}{|c|c|c|} \hline
  &   &   \\ \hline
 5 &  &  6 \\ \hline
 4 & 2 & 3 \\ \hline
\end{tabular}\hspace{2mm} U=4 , L=4}] {}
                edge from parent
                node[above] {}
                node[below]  {}
            }
            edge from parent
            node[above] {}
            node[below]  {}
    }
    child {
        node[bag] {U=5
        \vspace{2mm}
\\
\begin{tabular}{|c|c|c|} \hline
  &  6 &   \\ \hline
 1 & 5 &  \\ \hline
 4 & 2 & 3 \\ \hline
\end{tabular}
\vspace{2mm}
\\
L=3}
        child {
                node[end, label=right:
                    {\begin{tabular}{|c|c|c|} \hline
  &   &   \\ \hline
  & 5 & 6 \\ \hline
 4 & 2 & 3 \\ \hline
\end{tabular}\hspace{2mm} U=5 , L=4}] {}
                edge from parent
                node[above] {}
                node[below]  {}
            }
            child {
                node[end, label=right:
                    {\begin{tabular}{|c|c|c|} \hline
  &   &   \\ \hline
 6 & 5 &   \\ \hline
 4 & 2 & 3 \\ \hline
\end{tabular}\hspace{2mm} U=5 , L=4
}] {}
                edge from parent
                node[above] {}
                node[below]  {}
            }
        edge from parent
            node[above] {}
            node[below]  {}
    };
\end{tikzpicture}
\caption{Example of $A^*$ algorithm using H as upper bound and $S_0$ as lower bound (defined in subsection \ref{deflowerbounds})}
\label{SDTEx}
\end{figure}

\paragraph{Example 1}
Figure~\ref{SDTEx} presents the $A^*$ where the initial bay is shown in Figure~\ref{tab:ex}. It uses upper and lower bounds that are presented in Section~\ref{sec:bounds}. None of the nodes at level 1 can be pruned. Now let us consider the 4 nodes at level 2 (the 4 rightmost configurations in Figure~\ref{SDTEx}). The third one uses rule (i) to stop branching
from this node since lower and upper bounds are both equal to 4.
At this level $z_A=4$. Therefore we can also prune the first,
second and fourth nodes using rule (ii). Therefore, the tree is complete at
level 2. We follow the path until the third node at level 2, and then follow the
feasible solution given by the upper bound.
\\

We know from~\cite{RePEc} that the restricted CRP is NP hard. Therefore, there
exist some instances for which the number of nodes needed to find the optimal solution with this method grows exponentially with the problem size. However to limit the computation time, we set a maximum number of nodes $\mathcal{N}$
allowed in the $A^*$ tree. If the number of nodes reaches $\mathcal{N}$, we return our best upper-bound as a feasible solution.
In this case, we cannot insure optimality, but we can provide a guaranteed gap with the optimum solution.
Let $L_{min}$ be the minimum of the cumulative lower bounds of
every leaf of the tree that is not pruned by $A^*$. Notice that $z_{opt}(B) \ge L_{min}$. Thus
$z_A-z_{opt}(B) \le z_A-L_{min}=\gamma$ (line 24 of Algorithm~\ref{algo}).

We show later that given certain properties on the lower bounds, $\gamma$ is nonincreasing in $\mathcal{N}$. In any case, there exists $\mathcal{N}^*$ such that if $\mathcal{N}>\mathcal{N}^*$ then $\gamma=0$.
In our numerical average-case analysis (Section \ref{sec 5-2}), we show how $\mathcal{N}$ and $\gamma$ are related.
\\

The size of the $A^*$ tree depends on the local lower and upper bounds
used. The closer they are to the optimal solution at each node, the faster the tree is pruned,
hence the faster it solves. In this paper, we define some of the possible candidates, explain these choices
and provide details about the efficiency of the chosen bounds. Note that the algorithm can also be
implemented using depth first search instead of breadth first search; in this paper, we implement the latter as we find experimentally that it is the most efficient one on average.

\subsection{The Candidate Bounds}
\label{sec:bounds}
\subsubsection{The Lower Bounds}
\label{deflowerbounds}
\paragraph{The counting lower bound.}
This bound was introduced by Kim and Hong in~\cite{ref5} and it is based on the
following simple observation. In the initial configuration, if a container is blocking, then it must be
relocated at least once. Thus we count the number of blocking containers in $B$
and we denote it $S_0(B)$.
Note that if a container blocks more than one container (for instance, container 6 blocks containers 1 and 4 in
Figure~\ref{tab:ex}), we count this container only once.

\begin{fact}
  For any configuration B,
  $$ S_0(B) \le z_{opt}(B) $$
\end{fact}
  We define $L_0$ as the cumulative bound computed with $S_0$.
  \\
As mentioned in section~\ref{descalgo}, we want our algorithm to have the
following property: As we increase $\mathcal{N}$, $\gamma$ should not increase. One way to
enforce this is to insure that given any path of the tree, the cumulative lower bound is
nondecreasing on this path. We will indeed show that $L_0$ follows this property.
Before proceeding let us define a few notations: Let $c_i$ be the $i^{th}$
column of the configuration and $\min({c_i})$ be the index of the container with the smallest retrieval time in $c_i$.
Now suppose we want to move the blocking container $r$. If we relocate $r$ to column
$c_i$ and $r<\min({c_i})$, then we call such a move a ``good'' move and column $c_i$
is a ``good'' column for $r$. On the other hand, if $r>\min({c_i})$, we call this
move a ``bad'' move and $c_i$ a ``bad'' column for $r$.

\begin{prop}
\label{prop:path1}
  For any configuration B, any level $l \ge 0$, any configuration $B^l$ in level l of $A^*$ and
  any child $B^{l+1}$ of $B^l$, we have $L_0(B^{l+1}) \ge L_0(B^l)$.
\end{prop}


\paragraph{The look-ahead lower bounds.}
Note that the counting lower bound ($S_0$) is only taking into account the initial configuration of the bay. By
definition, a ``bad'' move implies a future relocation for the relocated
container. Therefore we can construct lower bounds
that anticipate ``bad'' moves and hence are closer to the optimal solution. This idea has
been used by Zhu et al (\cite{Astar1}) to define the following family of lower bounds. For the sake of completeness, we redefine the bounds formally. In this paper, we prove in
Proposition~\ref{prop:augmentingpathLN} that these lower bounds have the non-decreasing property
as we branch in the $A^*$ tree.
\\

The basic idea is the following: We want to count unavoidable ``bad'' moves in any feasible
sequence of moves. In order to do so, we need to keep track
of the ``maximum of minimums of each column of $B$'' denoted by $MM(B) = \underset{i \in \left\{ 1,\ldots,C \right\}}{\max} (min(c_i))$. Suppose $r$ is the first
blocking container; a ``bad'' move for $r$ is unavoidable if $MM(B) \le r$. As we empty the bay, the maximum of minimums is
changing. More precisely it depends on the sequence of moves that we use.
We can, however, compute an upper bound on this value for every feasible sequence of moves. One way is to
assume that every container that is relocated is discarded from the bay. At any time, the
maximum of minimums in the ``discarded'' bay will be not smaller than if we computed it given any feasible sequence of moves.
Before introducing formal notations, we give an example.
\paragraph{Example 2}
  \begin{figure}[htdp]
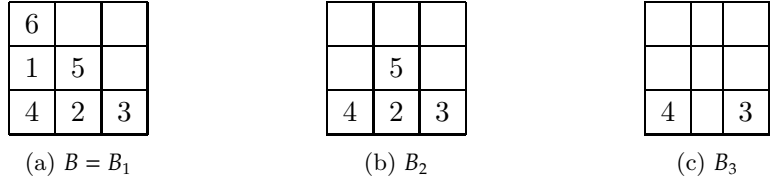

    \centering
\subfloat[$B=B_1$]{
\begin{tabular}{|c|c|c|} \hline
 6 &   &   \\ \hline
 1 & 5 &   \\ \hline
 4 & 2 & 3 \\ \hline
\end{tabular}
}
\hspace{20mm}
\subfloat[$B_2$]{
\centering
\begin{tabular}{|c|c|c|} \hline
    &   &   \\ \hline
    & 5 &   \\ \hline
 4 & 2 & 3 \\ \hline
\end{tabular}
}
\hspace{20mm}
\subfloat[$B_3$]{
\centering
\begin{tabular}{|c|c|c|} \hline
    &   &   \\ \hline
    &   &   \\ \hline
 4 &   & 3 \\ \hline
\end{tabular}
}
\caption{Example of discarded bays with 3 tiers, 3 columns and 6 initial containers}
\label{tab:exlowerbound}
\end{figure}

Consider the leftmost bay $B$ in Figure~\ref{tab:exlowerbound}. First, one can see that $S_0(B)=2$
since 6 is blocking 1, and 5 is blocking 2.
Now in $B_1$, 6 is blocking 1 and $MM(B_1)$ is 3; thus 6 has to be relocated at least twice.
Therefore the optimum is at least 1 more than $S_0$.
Now consider bay $B_2$ where 1 and 6 have been discarded. 5 is blocking 2 and $MM(B_2)$ is 4; thus 5 has
to be relocated at least twice. Therefore the optimum is at least 2 more than $S_0$.
\\

Now we formalize this idea: Let $n_1$ be the smallest container in
the bay. Let $k$ be a
container ($k \in \left\{n_1,\ldots,N\right\}$) and let $\mathcal{R}_k(B)$ be the set of containers blocking $k$ and
not blocking any container $k'$ such that $k' < k$.

Let $B_{n_1}=B$ and let $B_{n_1+1}$ be the bay where container $n_1$ and
containers in $\mathcal{R}_{n_1}(B)$ have all been discarded from $B_{n_1}$. By recursion, we can define
a sequence of bays $B_k$ for $k \in \left\{n_1+1,\ldots,N\right\}$.

For $p \in \mathbb{N}$, we now define the $p^{th}$ look-ahead lower bound (denoted by $S_p(B)$):
\begin{eqnarray}
S_p(B)=S_0(B)
+ \sum_{k=n_1}^{\min \left( p+n_1-1,N \right)} \sum_{r \in \mathcal{R}_k(B)} \chi(r>MM(B_k)),
\end{eqnarray}
where $\chi(\cdot)$ is the indicator function. The corresponding cumulative lower bound is denoted by $L_p(B)$.

\begin{fact}
  For every configuration $B$ and $p \in {1,\ldots,N}$, $$S_p(B) \le z_{opt}(B)$$
\end{fact}

\begin{prop}
  For $p \in {1,\ldots,N}$ and any bay B, we have $L_{p}(B) \ge L_{p-1}(B)$.
  \label{prop:LB decreasing}
\end{prop}

This result shows that as we increase $p$, $S_p$ gets closer to the optimal solution. In
Section \ref{sec 5-2}, we numerically study the effect of the choice of lower bounds in the $A^*$
algorithm in terms of the number of nodes needed to solve a random set of instances.
We observe that on average, most of the gain comes from the first look ahead.

We remark that $S_N(B)=S_{N-C}(B)$ simply because when we reach N-C, there is
always at least one empty column. Note that if $k$ is the smallest integer such
that $B_k$ has an empty column, then the lower bound does not change, i.e.,
$S_N(B)=S_{k-1}(B)$. Thus, in practice, the process of computing the lower bounds
terminates much before $N-C$.

In the next proposition, we show that similar to $L_0$, the cumulative upper bound $L_p$ has the desirable monotone property, i.e.,
they do not decrease as we branch in the $A^*$ tree.
\begin{prop}
\label{prop:path2}
  For any $p \in {1,\ldots,N}$, configuration B, level $l \ge 0$, configuration $B^l$ in
  level l and any child $B^{l+1}$ of $B^l$ in the $A^*$ tree, we have $L_p(B^{l+1}) \ge L_p(B^{l})$.
\label{prop:augmentingpathLN}
\end{prop}

In general it is hard to find the rate at which the lower bound increases on a given
path. In Section \ref{sec 5-2}, we show how the lower bound $S_N$ increases on the optimal
path of random instances.

\subsubsection{The Upper Bound}
Any feasible solution can be used as an upper bound. Thus, we use heuristics to
construct feasible solutions that serve as upper bounds.

Ideally we would want a heuristic that is close to the optimal solution and at
the same time easy to compute. Note that we need to compute an upper bound at every
node. The heuristic proposed by Casserta et al. (\cite{RePEc}) seems to meet
those criteria. For completeness, we redefine this heuristic. In this article we
prove several properties of the upper bound that we will use in later parts
(see Propositions~\ref{prop:caseCcontainers},
~\ref{prop:caseC1containers} and~\ref{prop:caseCkcontainers}).

\paragraph{The Heuristic H (\cite{RePEc})}
Suppose $n$ is the target container located in column $c$, and $r$ is the topmost blocking
container in $c$. We use the following rule to determine $c^*$, the column where $r$ should be relocated
to. Recall that $\min (c_i)$ is the minimum of column $c_i$. We set $\min (c_i)=N+1$ if $c_i$ is empty.
We have:
\[c^* = \left\{
\begin{array}{l l}
 \underset{c_i \in \{ 1,\ldots,C \} \setminus c}{\text{argmin}} \lbrace{\min (c_i):\min (c_i)>r}\rbrace & \quad \text{if
 $\exists$ $c_i$ such that $\min (c_i)>r$} \\ \underset{c_i \in \{ 1,\ldots,C \} \setminus c}{\text{argmax}} \lbrace{
 \min (c_i) }\rbrace & \quad \text{otherwise}\\ \end{array} \right. \]

The rule says: if there is a column where $\min (c_i)$ is greater than $r$ ($r$ can do a ``good'' move), then
choose such a column where $\min (c_i)$ is minimized, since
columns with larger minimums can be useful for larger blocking containers.
If there is no column satisfying $\min (c_i) > r$ ($r$ can only do ``bad'' moves), then choose the column where
$\min (c_i)$ is maximized in order to delay the next unavoidable relocation of $r$ as much as possible.
 We will refer to this heuristic as heuristic H and denote its number of relocations by $z_{H} (B)$.

 \paragraph{Example 2}
Consider the bay $B$ in Figure~\ref{tab:ex}. Using heuristic H, the sequence is
going to be: relocation of 6 to column $c_3$, retrieval of 1, relocation of 5 to column $c_1$, retrieval of 2,
relocation of 6 to column $c_2$, retrieval of 3, relocation of 5 to column $c_2$, retrievals of 4,5 and 6. Thus
$z_H(B)=4$. Notice that we had $S_2(B)=4$, so heuristic H is optimal
for this initial configuration.
\\

 By definition, H is a feasible solution, so
 it is an upper bound on the optimal solution.
\begin{fact}
  For any configuration B, we have $z_{opt} (B) \le z_{H} (B)$.
  \end{fact}

The heuristic H has a certain number of useful properties stated below.
\begin{prop}
In a bay with $C$ columns, for any configuration $B$ with at most $C$ containers, we
have
  \begin{eqnarray}
\label{caseCcontainers}
 S_0(B) =  z_{opt} (B) = z_{H} (B).
  \end{eqnarray}
\label{prop:caseCcontainers}
\end{prop}

\begin{prop}
  In a bay with $C$ columns, for any configuration $B$ with at most $C+1$ containers, we
have $S_1(B) = z_{opt} (B) = z_{H} (B)$.
\label{prop:caseC1containers}
\end{prop}

This result implies that if we use heuristic H as an upper bound
and the $S_1$ lower bound in the $A^*$ approach, the tree will stop at most after $N-C-1$ retrievals with
the guarantee of an optimal solution.

We provide two similar bounding results on H.
\begin{prop}
 In a bay with $C$ columns, for any configuration $B$ with at most $C+k$ containers, we have:
 \begin{itemize}
   \item $z_H(B) \le z_{opt}(B)+2$, if $k = 2$
   \item $z_H(B) \le z_{opt}(B)+\frac{k(k+1)}{2}$, if $3 \le k \le C$.
 \end{itemize}
\label{prop:caseCkcontainers}
\end{prop}

\begin{remark}
  The following example shows that this upper bound can increase as we branch in the $A^*$ tree.
  In Figure~\ref{tab:exupperbound}, bay (a)
  shows the initial configuration and bay (b) is one of its two ``children''. It is
  easy to check that heuristic H needs 7 relocations for bay (a) and 8 for bay (b).

  \begin{figure}[htdp]
    \centering
\subfloat[]{
\begin{tabular}{|c|c|c|} \hline
    &    &    \\ \hline
 5 & 9 & 7 \\ \hline
 1 & 2 & 4 \\ \hline
 3 & 6 & 8 \\ \hline
\end{tabular}
}
\hspace{20mm}
\subfloat[]{
\centering
\begin{tabular}{|c|c|c|} \hline
    & 5 &    \\ \hline
    & 9 & 7 \\ \hline
    & 2 & 4 \\ \hline
 3 & 6 & 8 \\ \hline
\end{tabular}
}
\caption{Heuristic H can increase on a path of $A^*$ tree}
\label{tab:exupperbound}
\end{figure}
\end{remark}


\section{An asymptotic analysis of CRP}
\label{sec:asym}
In this section, we study CRP for random large bays and we show that as the
number of columns in the bay grows, the problem gets ``easier'' in the sense
that the gap between the optimal solution and our simplest lower bound ($S_0$),
does not increase on average, and in fact it is bounded by a constant. The basic intuition is that,
as the number of columns grows, for any blocking container, we can find a ``good'' column with high
probability. This implies that each blocking
container is only relocated once with high probability.

Throughout this section, we assume that initial configurations have $P$ tiers, $C$ columns,
$N$ containers and that there are exactly $h$ containers in each column, where $h \le P-1$. For more clarity we denote by $B_C$ such a bay with $C$ columns.We assume
that the initial configuration is a uniform sample among all possible such
configurations.
 Notice that in that case, when $C$ grows to infinity, $N=h \times C$ also grows to infinity.
 \\

First let us explain how a uniformly random bay is generated.
We view a bay as an array of $P \times C$ columns. The slots are numbered from bottom to top, and left to right from 1 to $PC$.
For example, let $P=4$ and $C=7$, the second topmost slot of the
third column is 11.
The goal is to get a bay $B$ with uniform probability, meaning each container is equally likely to be
anywhere in the configuration, with the restriction that there are $h$ containers per column.
We first generate uniformly at random
a permutation of $\{1,\ldots,N\}$ called $\pi$. Then we assign a slot for each container
with the following relation: $B(j,i)=\pi(h\times (i-1) + j)$ for $j \le h$ and $B(j,i)=0$
for $j \ge h+1$. One can see that each bay is generated with probability $\frac{1}{N!}$.
 There is a one to one mapping
between configurations with $C$ columns and permutations of $\{1,\ldots,hC\}$, denoted by $\mathcal{S}_{hC}$.
Finally, we denote the expectation of random variable X over
this uniform distribution by $\mathbb{E}_{C}[X]$.
\\

Now we compute the expected counting lower bound.
 \begin{prop}
\label{lemmalinlowerbound}
   Let $S_0$ be the counting lower bound (defined in Section \ref{sec:SDT}), we have
   \begin{eqnarray}
     \mathbb{E}_C\left[ S_0 (B_C)\right] = \alpha_h \times C,
   \end{eqnarray}
   where $\alpha_h$ is the expected number of
   blocking containers in one column and can be computed as
   \begin{eqnarray}
     \alpha_h= \sum_{k=1}^{h-1} k \times p_{k,h}
   \end{eqnarray}
   and $p_{k,h}=\mathbb{P}[\text{there are k blocking containers in a column with h
  containers}]$ can be computed by recrusion as
  \begin{eqnarray*}
  \forall h \ge 0 \textit{, } p_{0,h} = \frac{1}{h!} \textit{ and } \forall k \ge 1\textit{, }  p_{k,h}=\sum_{j=1}^k \frac{1}{h} p_{k-j+1,h-j} \
  \end{eqnarray*}
 \end{prop}

 \begin{remark}
\label{remarkalpha}
   Note that $\alpha_h$ only depends on the distribution of the relative order of the $h$ containers in one column (and not on the actual labels).  Thus the expected number of blocking containers in one column only depends on its height.
 \end{remark}

 The major result of this part is the following. In the asymptotic case where
 the number of columns increases to infinity, the expected optimal number of relocations is asymptotically proportional to the expected number of blocking containers (the counting lower bound).

 \begin{theorem}
 \label{theor:asym}
  Let $S_0$ be the counting lower bound (defined in Section \ref{sec:SDT}) and $z_{opt}$ be the optimal
  number of relocations. Then for $C \ge h+1$, we have
   \begin{eqnarray}
     1 \le
     \frac{\mathbb{E}_C\left[ z_{opt} (B_C)\right]}{\mathbb{E}_C\left[ S_0 (B_C)\right]}
      \le f(C)
   \end{eqnarray}
   where
   \begin{eqnarray}
     f(C)=1+\frac{K}{C}
     \underset{C \rightarrow\ \infty}{\rightarrow} 1
   \end{eqnarray}
   where K is a constant defined by equation (\ref{constantK}).
\end{theorem}

The proof of Theorem \ref{theor:asym} is given in the Appendix. Here we just give an intuition of the proof. We show that as $C$ grows, with high probability $\mathbb{E}_{C+1}[z_{opt}(B_{C+1})]-\mathbb{E}_{C}\left[z_{opt}(B_C)\right]$ is exactly $\alpha_{h}$. Therefore, for $C$ large enough, $\mathbb{E}_{C}\left[z_{opt}(B_C)\right]$ essentially behaves like $\alpha_{h}\times C$, which is equal to $\mathbb{E}_C[S_0(B_C)]$ (according to Proposition~\ref{lemmalinlowerbound}).

In the next corollary, we show that the optimal solution of the unrestricted CRP has a similar asymptotic behavior. We remind that the unrestricted CRP refers to the problem where we can also relocate non-blocking containers.
\begin{corollary}
  Let $z_{gen} (B_C)$ be the optimal number of relocations for the unrestricted C\@R\@P.
 For $C \ge h+1$, we have
\begin{align}
  1 \le \frac{\mathbb{E}_C\left[ z_{gen} (B_C)\right]}{\alpha_h C} \le f(C)
  \label{equival2}
  \end{align}
  where f is the function defined in Theorem~\ref{theor:asym}.
 \end{corollary}

\begin{figure}[h]
\centering
\subfloat[Convergence of the ratio $\frac{\mathbb{E}\left[ z_H (B_C)\right]}{\mathbb{E}\left[ S_0 (B_C)\right]}$]
{\includegraphics[width=0.4 \textwidth]{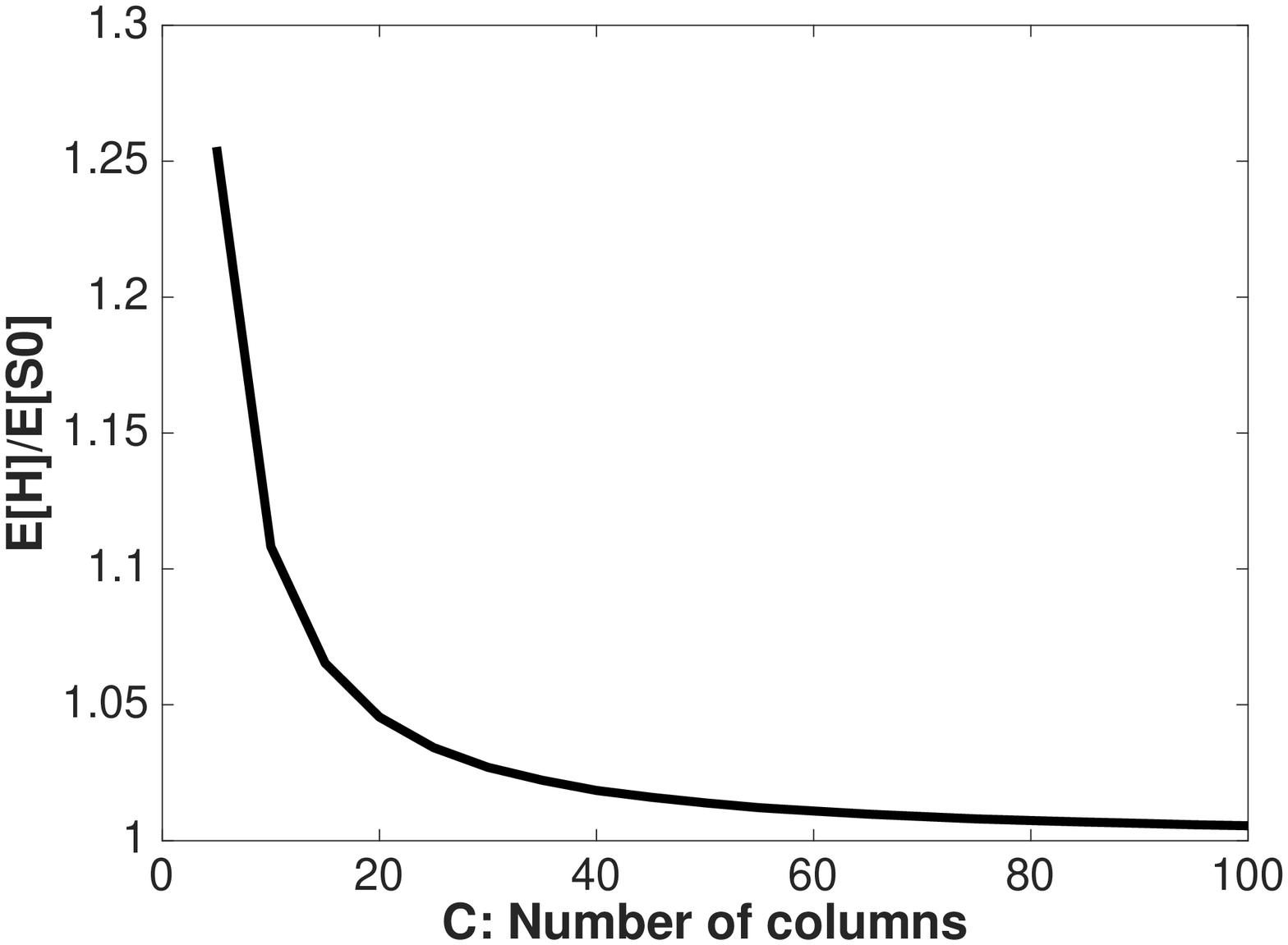}
\label{fig:Exp4a}}
\quad \hspace{1cm}
\subfloat[Convergence of the difference ${\mathbb{E}\left[ z_H (B_C)\right]} - {\mathbb{E}\left[ S_0 (B_C)\right]}$]
{\includegraphics[width=0.4 \textwidth]{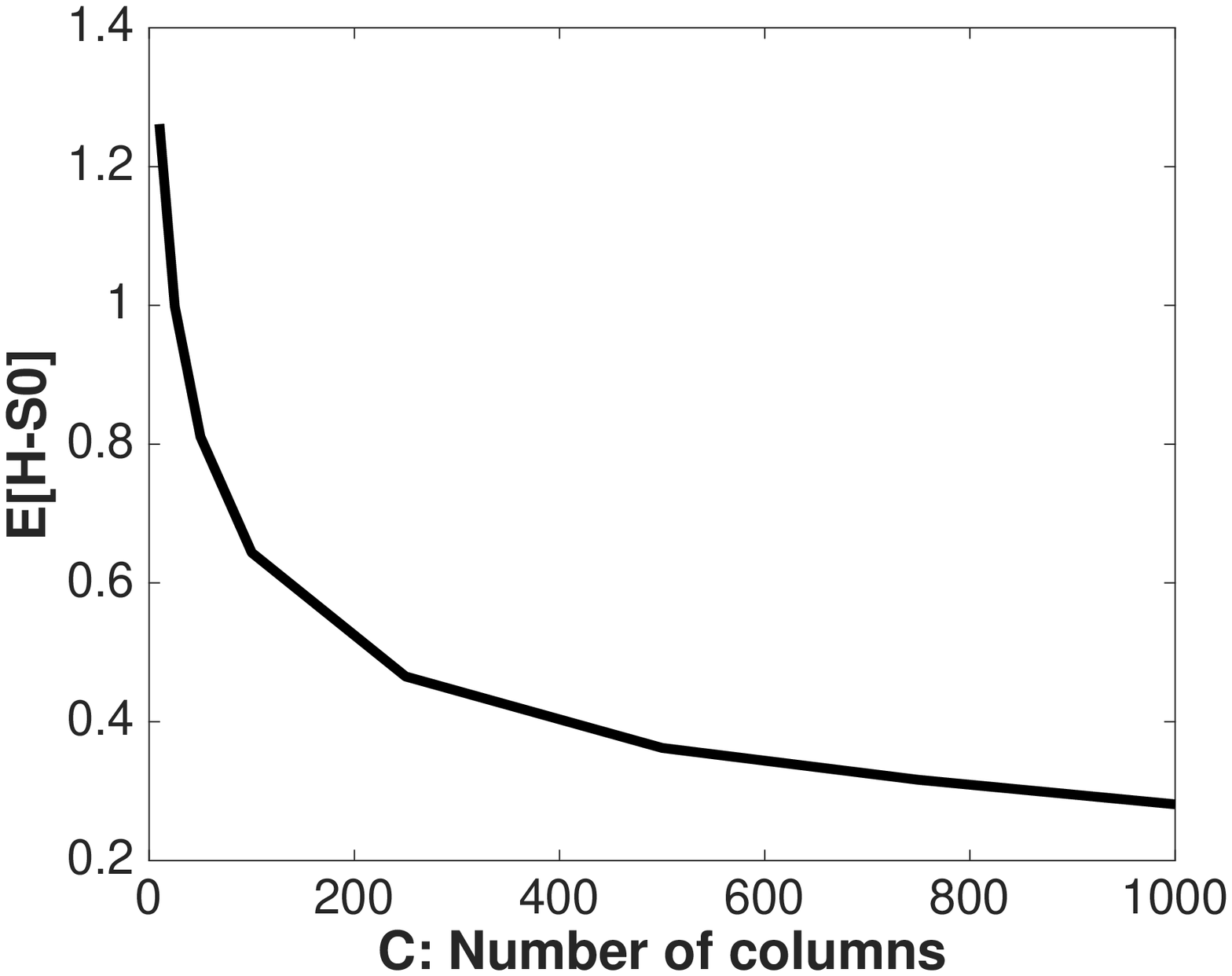}
\label{fig:Exp4b}}
\caption{Simulations for the Asymptotic Theorem}
\end{figure}

The above theorem gives insights on how the expected optimal
solution of the CRP behaves asymptotically on random bays. To conclude this section, we show experimentally that the same result holds for heuristic H, i.e., the ratio of $\mathbb{E}\left[ z_H (B_C)\right]$ and $\mathbb{E}\left[ S_0 (B_C)\right]$ converges to 1 as $C$ goes to infinity.
For each size C, we compute both expectations
over a million instances, take their ratio and plot the result in Figure~\ref{fig:Exp4a}. Notice that we have
$1 \le \frac{\mathbb{E}\left[ z_{opt} (B_C)\right]}{\mathbb{E}\left[ S_0(B_C)\right]}
\le \frac{\mathbb{E}\left[ z_H (B_C)\right]}{\mathbb{E}\left[ S_0 (B_C)\right]}$, so Figure~\ref{fig:Exp4a} also
shows experimentally that Theorem \ref{theor:asym} holds.

First, note that Figure \ref{fig:Exp4a} implies that the relative gap between
heuristic H and $S_0$ shrinks to 0 as C increases. Moreover we have
$\frac{\mathbb{E}\left[ z_H (B_C)\right]-\mathbb{E}\left[ z_{opt} (B_C)\right]}{\mathbb{E}\left[ z_{opt} (B_C)\right]}
\le
\frac{\mathbb{E}\left[ z_H (B_C)\right]-\mathbb{E}\left[ S_0 (B_C)\right]}{\mathbb{E}\left[ S_0
(B_C)\right]}$ and thus the relative gap of H with optimality also converges to 0 as C
grows to infinity.

In the proof of Theorem~\ref{theor:asym}, we also study the function $g(C)=\mathbb{E}_C\left[ z_{opt} (B_C)\right]
-\mathbb{E}_C\left[ S_0 (B_C)\right]$.

Note that $g(C) \le \mathbb{E}_C\left[ z_{H} (B_C)\right]-\mathbb{E}_C\left[ S_0
(B_C)\right]$ where the right-hand side of the inequality is the function plotted in
Figure~\ref{fig:Exp4b}. The plot shows that $g(C) \le 1.25$ for all C, meaning that $g(C)$ is bounded as we proved in Theorem \ref{theor:asym}.
Moreover, the plot implies that heuristic H is on average at most 1.25 away from the optimal solution, so heuristic H is relatively more efficient
in the case of large bays. Intuitively,
the probability of having a good column converges to 1, as we increase the number of columns; hence the problem tends
to become easier as C grows.

Finally, in the proof, we note that the rate of convergence for the optimal solution to $S_0$
is at least $\frac{1}{C}$. Interestingly, we can infer from Figure \ref{fig:Exp4a}.
that the rate of convergence of the ratio for heuristic H is also proportional to $\frac{1}{C}$.

\section{CRP with Incomplete Information}
\label{sec:intro}
The $A^*$ algorithm as explained in Section \ref{sec:SDT}, relies on the assumption that the departure order of containers is known in advance. However, in practice, we often only know the departure order of some of the containers (those that are going to depart in the near future).

In this section, we explain how the $A^*$ algorithm can be adapted in a Dynamic Programming (DP) framework for solving the CRP in the case that some of the decisions should be made with incomplete information.
We use the $A^*$ algorithm to obtain an approximate solution for this problem. In what follows, we describe the setting of the problem, introduce some notations, and explain the algorithm for the CRP with incomplete information. Moreover, we show how to bound the approximation error. In Section \ref{sec 5-2}, we present the results of computational experiments.

The CRP with incomplete information involves retrieving $N$ containers from a bay with $C$ columns and $P$ tiers, where partial information about the departure order of the containers is initially available.
To model the CRP with incomplete information, we discretize time into time steps of equal length and assume that each move (a relocation or a retrieval) takes exactly one time step.
Further we assume that the label (or index) of a container indicates the earliest time that it can be retrieved (i.e., container $n$ can be retrieved at time $n$ or later).
Usually, very little information is available about the containers that are going to be retrieved far in the future. Thus, it is reasonable to assume that at any given time step $t$, we only know the departure order of a certain number of containers in the bay (the containers that are going to be retrieved within a short time horizon after $t$). We refer to such containers in the bay as \emph{known} containers. Similarly, we refer to the remaining containers in the bay as \emph{unknown} containers.
By definition, all unknown containers have larger indices (i.e., later departure times) than the known containers.


As time passes, some of the known containers are retrieved and as more information becomes available, some of the unknown containers become known. In general, the information might be updated multiple times during the retrieval process (every time that the arrival time of a truck is provided by the truck driver, the information gets updated and some unknown containers become known). In the most general case, the information can be updated every time step.
Alternatively, we can consolidate several small pieces of information into one or a few pieces, and assume that the information is revealed at $\Gamma$ different times during the retrieval process (i.e., $\Gamma$ is the number of time that information is updated). In this case, we have a multi-stage problem and $\Gamma$ sets of decisions need to be made.

In this paper, we focus on a 2-stage setting; we assume that a subset of containers is initially known and that the departure order of all the remaining ones becomes known at time $t^*$. We denote the set of known containers by $\mathbf{K}$ and assume that containers $\{1, 2, \dots, |\mathbf{K}|\}$ are those known at time zero. Similarly, the set of unknown containers is denoted by $\mathbf{U}$, and containers $\{|\mathbf{K}|+1, \dots,N\}$ become known at time step $t^{*}>0$. We refer to this setting as the \emph{2-stage} CRP since there are two types of decisions that need to be made: first-stage decisions (retrievals and relocations before time step $t^{*}$) and second-stage decisions (retrievals and relocations after time step $t^{*}$). We assume that before $t^*$, we have probabilistic information about the containers in $\mathbf{U}$, meaning that we know the probability of realization of each possible departure order. Such information can be obtained from historical data or from an appointment system that provides some estimate of departure times of the containers (for example a time window for retrieving each container). We denote the set of possible departure orders of the containers in $\mathbf{U}$ (possible scenarios) by $\mathcal{Q}$. From now on, we assume that all scenarios are equally likely, i.e., the probability of each scenario is $\dfrac{1}{|\mathcal{Q}|}$, and the number of scenarios is $(N - |\mathbf{K}|)!$. Note, however, that we could use the algorithm to solve the CRP with any other probability distribution on the departure order of unknown containers.

We use a 2-stage stochastic optimization technique to solve this problem, where in the first-stage we minimize $\mathbb{E}[z]$ as follows:

\begin{eqnarray}
\min_{\sigma_1, \dots, \sigma_{t^*-1}} && \mathbb{E}[z] = \sum\limits_{q\in \mathcal{Q}} \dfrac{1}{|\mathcal{Q}|} z(B(q)).
\label{obj_2stage}
\end{eqnarray}

where $B(q)$ is the resulting bay when scenario $q$ is realized, $z(B(q))$ is the total number of relocations for $B(q)$, and $\sigma_1, \dots, \sigma_{t^*-1}$ are the first-stage decisions. The 2-stage problem can be solved with the $A^*$ algorithm as follows:

\textbf{(1).} We build the tree with $t^* - 1$ moves for time-steps $1,2,\dots,t^*-1$, in a similar way as illustrated in Figure \ref{SDTEx}; we denote this tree by $T_{[1,t^*-1]}$.

\textbf{(2).} For each node at level $t^*-1$, we need to compute the expected number of remaining relocations. We enumerate all possible scenarios and solve the CRP with complete information corresponding to each scenario, using the $A^*$ algorithm.

\textbf{(3).} We find $p^*$ (the optimal path or sequence of moves) that minimizes the expected total number of relocations over all paths.

\textbf{(4).} Once $p^*$ up to time-step $t^*-1$ is selected and we observe the information at $t^*$, we use the $A^*$ algorithm to solve a specific instance through the end.

Notice that to find the optimal path for time interval $[1,t^*-1]$, we need to solve up to $(C-1)^{(t^*-1)}(N-|\mathbf{K}|)!$ instances in Step (2) with the $A^*$ algorithm. Although $A^*$ is fast, the number of scenarios is prohibitively large and enumerating all scenarios is not feasible due to limited resources of memory and long computation time. We next explain how we use \emph{sampling} and \emph{pruning} to overcome these issues. We also quantify the error incurred as a result of sampling and pruning. We refer to the resulting algorithm as $ASA^*$ (Approximate Stochastic $A^*$).

\textit{\textbf{Limiting the number of possible scenarios on each path.}} We overcome this issue by sampling $S$ scenarios on each path and computing the number of relocations for the sampled departure orders rather than for all possible orders. For each path $p$, let $\bar{z}$ be the number of relocations averaged over samples. Also let $\mathbb{E}[z]$ be the true mean. To determine the number of samples needed to get a good approximation, we use the following version of Hoeffding's inequality:

\begin{eqnarray}
P(|\mathbb{E}[z] - \bar{z}| > \delta ) \le 2\exp\left(\dfrac{-2S\delta^2}{(r_{max}-r_{min})^2}\right),
\label{hoeffding}
\end{eqnarray}

where $\delta$ is a pre-specified level of desired precision, and $r_{max} \text {/} r_{min}$ are lower/upper bounds on random variable $z$.

Note that we can set $r_{min}=0$; for $r_{max}$, we do not have a tight bound, but we can use $N(P-1)$ as an upper bound (since each of the $N$ containers is blocked by at most $P-1$ containers). Let us denote the desired probability for bounding the error (i.e., RHS of \eqref{hoeffding}) by $\epsilon$. For a given $\delta$ and $\epsilon$, the required number of samples can be computed as follows.

\begin{eqnarray}
S \ge \dfrac{{r^2}_{max} \ln\frac{\epsilon}{2}}{-2\delta^2}.
\label{numSample}
\end{eqnarray}

By sampling from the possible scenarios on each path, we can significantly reduce the number of scenarios (and thus the computation time). For example, for a bay with 7 columns, 4 tiers, 21 containers, and $|\mathbf{K}|=6$, the total number of possible scenarios on any of the paths at time $t^*$ is about $10^{11}$. Using inequality \eqref{numSample}, the total number of scenarios would be around 30,000 for $\delta=0.5$, $\epsilon=0.05$, and $r_{max}=63$. Note that by sampling, we incur an error and $ASA^*$ may choose a suboptimal path $p_{ASA}$ where $\mathbb{E}[z_{p_{ASA}}] > \mathbb{E}[z_{p^*}]$ (recall that $p^*$ is the optimal path that would be chosen without sampling). In the next proposition, we show that such an error, denoted by $e_1 \overset{\Delta}{=} \mathbb{E} [z_{p_{ASA}}]  - \mathbb{E} [z_{p^*}]  $, is bounded in expectation (over a uniform distribution on the initial bay).

\begin{proposition}
Suppose for each path, we estimate the number of relocations using $S$ independent samples, where $S$ is given in \eqref{numSample}.
\textcolor{black}{Also suppose $ASA^*$ chooses path $p_{ASA}$ as the optimal path, and $e_1 = \mathbb{E} [z_{p_{ASA}}]  - \mathbb{E} [z_{p^*}]  $. We have  $\mathbb{E} [e_1] \le 2\delta \sqrt{\dfrac{\pi}{-\ln({\dfrac{\epsilon}{2}})}}$.}
\label{pro1}
\end{proposition}

\begin{figure}[htb!]
 \centering
 {\includegraphics[width=3cm,height=6cm]{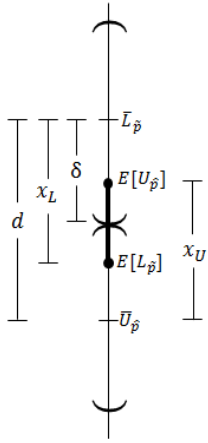}}
 \caption{Illustration of the error due to pruning: the thick line segment is the error as a result of pruning the optimal path by mistake.}
 \label{ublb}
 \end{figure}

\textit{\textbf{Pruning the paths of $T_{[1,t^*-1]}$.}} To address this issue, we use the upper bound and lower bounds to prune the nodes of $T_{[1,t^*-1]}$, similar to the $A^*$ algorithm. However, since some of the containers are unknown before $t^*$, we have to compute the expectations ($\mathbb{E}[{L}]$ and $\mathbb{E}[{U}]$). Again, we use the idea of sampling and we estimate these values by computing $\overline{L}$ and $\overline{U}$ using $S$ samples, where $S$ is obtained from Inequality \eqref{numSample}. Because of the sampling error, we may prune an optimal path by mistake, resulting in an error that is illustrated in Figure \ref{ublb}. This error is the difference between $\mathbb{E}(U_{\hat{p}})$ (the true upper bound of the path with the minimum estimated upper bound), and $\mathbb{E}(L_{\tilde{p}})$ (the true lower bound of the path selected by $ASA^*$ for pruning). Intuitively, this is because $\mathbb{E}(U_{\tilde{p}})$ is the worst we would achieve if we prune $\tilde{p}$ (by mistake), and  $\mathbb{E}(L_{\hat{p}})$ is the best we could achieve if we do not prune $\tilde{p}$.

In the next two propositions, we show that the error due to pruning is bounded, whether we prune some paths once at $t^*-1$ (resulting in error $e_2$) or prune some paths at several time-steps before $t^*-1$ (resulting in error $e_3$).

\begin{proposition}
Suppose that for each path $p$ at time-step $t^*-1$, we estimate the expected lower and upper bounds ($\overline{L}_{p}$ and $\overline{U}_{p}$) from $S$ samples, where $S$ is given in \eqref{numSample}. Also, suppose $ASA^*$ chooses to prune one or more paths.
We have $\mathbb{E}[e_2] \le 2\delta \sqrt{\dfrac{\pi}{-\ln({\dfrac{\epsilon}{2}})}}$.
\label{pro2}
\end{proposition}

\begin{proposition}
Suppose we prune some paths at $m$ time-steps $t_1, t_2, ..., t_m <t^*-1$. At each time of pruning and for each path $p$, we estimate the expected lower and upper bounds ($\overline{L}_{p}$ and $\overline{U}_{p}$) from $S$ samples where $S$ is given in \eqref{numSample}. Also, suppose that at each time, $i$, of pruning, $ASA^*$ chooses to prune one or more paths (denoted by $\tilde{p}_i$).
The expected total loss, $\mathbb{E}[e_3]$, is bounded by
$m \textrm{ }\left[
\textrm{ } (\dfrac{\epsilon}{2})^{\frac{d_{min}^2}{\delta^2}} +
\dfrac{d_{min}}{\delta} \sqrt{\dfrac{-ln(\dfrac{\epsilon}{2})\pi}{2}}\textrm{ } (\dfrac{\epsilon}{2})^{\frac{d_{min}^2}{2\delta^2}}
\right]
\left(
\delta \sqrt{\dfrac{\pi}{-\ln({\dfrac{\epsilon}{2}})}} + \overline{U}_{\hat{p}}^{max}
\right)$
, where
$\overline{U}_{\hat{p}}^{max} = \max\{\overline{U}_{\hat{p}_1},\dots,\overline{U}_{\hat{p}_m}\}$,
$d_i = \overline{L}_{{\tilde{p}}_i} -\overline{U}_{\hat{p}_i}$,
$d_{min} = \min\{d_1,\dots,d_m\}$.
\label{pro3}

\end{proposition}

In Propositions \ref{pro1}-\ref{pro3}, we bound the loss that can be incurred by pruning some paths at time $t^*-1$ $(e_1$ and $e_2)$ or at $t<t^*-1$ ($e_3$).
Notice that  for fixed $\delta$ and $\epsilon$, the errors  $e_1$ and $e_2$ are independent of the bay size if we increase the number of samples $S$ according to \eqref{numSample};
thus if we prune paths only at $t^*-1$, the loss remains unchanged and the relative loss (as a percentage of total relocations) decreases as the bay gets larger. Therefore, for large bays and ``hard-to-solve'' configurations, one can pick larger $\delta$ and $\epsilon$ that result in a smaller number of samples, and the relative error would still be small.

Table \ref{table_error} shows $e_1$, $e_2$ and $e_3$.
The losses are very small even for combinations of $\delta$ and $\epsilon$ that result in reasonably small number of samples. For example, for $\delta\le 1$ and $\epsilon < 0.1$, $\mathbb{E}[e_1]$ and $\mathbb{E}[e_2]$ are no more than 2. Because $e_1$ and $e_2$ are independent of the bay size, the relative error ($\dfrac{e_1+e_2}{\mathbb{E}[z]}$) decrease as the bay gets larger ($\mathbb{E}[z]$ is the average number of relocations when full information is available). For the $e_3$, this measure is almost constant for different bay sizes and can be controlled by changing $\epsilon$, $\delta$, and $m$ (number of times that we do pruning at $t<t^*-1$),

\begin{table}[htb]
 \begin{center}
 \begin{tabular}{ c c c}
  \hline
 $\delta$ &  $\epsilon$ & $\mathbb{E}[e_1]$ and $\mathbb{E}[e_2]$ \\
 \hline
0.1 & 0.01 &  0.15 \\
0.1 & 0.05 &  0.18  \\
0.1 & 0.1 &  0.2  \\
0.5 & 0.01 & 0.77  \\
0.5 & 0.05 & 0.92  \\
0.5 & 0.1 &  1.02  \\
1 & 0.01 &  1.54  \\
1 & 0.05 & 1.85  \\
1 & 0.1 &  2.05  \\
\hline
\end{tabular}
\quad\quad\quad
 \begin{tabular}{ c c c}
  \hline
 $C$  & $\mathbb{E}[e_3]$ & $\%$ error($\frac{\mathbb{E}[e_3]}{\mathbb{E}[z]}$) \\
 \hline
10 & 0.91 & 0.07\\
15 & 1.36 & 0.073\\
20 & 1.81 & 0.074\\
25 & 2.26 & 0.075\\
30 & 2.72 & 0.076\\
35 & 3.17 & 0.076\\
40 & 3.62 & 0.076\\
45 & 4.07 & 0.076\\
50 & 4.52 & 0.076\\
\hline
\end{tabular}
\end{center}
\caption{Left table: Expected loss due to sampling and pruning at $t^*-1$, for $C=7$, $P=4$, and $N=21$;
Right table: Expected loss due to pruning at $t<t^*-1$ for $\epsilon=0.05$, $\delta=0.5$,
$d_{min}=1$, $\overline{U}_{\hat{p}}^{max}=2N$, and $m=5$ }
\label{table_error}
\end{table}

In Section \ref{sec 5-2}, we use $ASA^*$ to solve the CRP with incomplete information for a bay with 7 columns and 4 tiers and with different amounts of information initially available. Moreover, we introduce a heuristic for this problem and compare the results of $ASA^*$ with those of the heuristic.

\section{Experimental Results}
\label{sec 5-2}
In this section, several experimental results are presented to further understand the effectiveness of our
algorithms in both complete and incomplete information settings.

In the complete information case, we study thoroughly the $A^*$ algorithm and the effect of parameters
on the efficiency and performance of the algorithm. First we show that using the $N^{th}$ look-ahead lower bound improves
dramatically the computational tractability of the algorithm. Moreover, our experiments shows that on the optimal path of the tree, the $N^{th}$ look-ahead lower bound reaches the optimum solution after a few levels.
We also show the trade-off between
$\mathcal{N}$ (maximum number of nodes in the tree) and the number of instances solved optimally.
Further we introduce a new class of heuristics, the Tree Heuristic (T\@H-L), compare it with 3 existing heuristics (\cite{ref5}, \cite{RePEc}, \cite{ref9}) and show that it outperforms the existing heuristics.

In the incomplete information setting, we show through experiments that the $ASA^*$ algorithm is fast and efficient and most of the instances are solvable within a reasonable time for medium-sized instances. We introduce a myopic heuristic which expands
on the H heuristic in \cite{RePEc}. Using the $ASA^*$ algorithm and the heuristic, we study
the value of information. Our simulations results show that while more information results in fewer relocations, the marginal value of information decreases with the level of information. We also use the myopic heuristic to examine the
effect of the level of information on the number of relocations for different bay sizes. We show that for any level of information, the ratio of relocations with incomplete information and complete information converges to a constant as the bay gets larger.

\subsection{Experimental results for complete information}
 In most of the experiments, we consider medium-sized instances of the CRP.
 We consider bays of size 4 tiers and 7 columns with 21 containers ($P=4=h+1$, $C=7$ and $N=21=hC$
 with 3 containers per column). We generate randomly 100,000 instances using
 the uniform distribution on bays described in Section 4.

\paragraph{The effect of lower bounds on the size of the $A^*$ Tree.}
In Section~\ref{sec:SDT}, we introduced a family of lower bounds ($S_0$,
$S_1$,\ldots,$S_N$). Clearly $S_N$ is more computationally expensive to use than
$S_0$, but we proved that $S_N$ was a tighter lower bound. We show here that using $S_N$ improves the performance of the $A^*$ algorithm significantly.

\begin{figure}[h]
\centering
\includegraphics[width=0.5\textwidth]{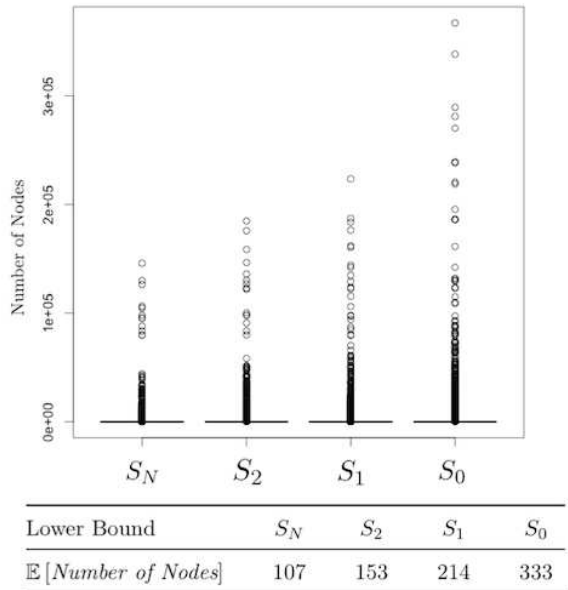}
\caption{The effect of the Lower Bound on $A^*$}
\label{fig:Exp1}
\end{figure}

For each instance, we solve it using 4 types of lower bounds $S_0$, $S_1$, $S_2$
and $S_{21}$ (here $N=21$). For each of them we record the number of nodes needed to
solve the instance optimally. Figure~\ref{fig:Exp1} presents the results. First, we show in the box-plots,
the distribution of the number of nodes for each of those lower bounds.
Second, we give the average number of nodes.
We give two main insights from this experiment. First, the average number of nodes
needed to solve with $S_{21}$ is 1/3 of the number of nodes with $S_0$. Also, we point out that
using $S_1$ and $S_2$ instead of $S_0$ decreases the average number of nodes by 1/3 and 1/2, respectively.
The fact that the number of nodes is decreasing as we use tighter lower bounds is not surprising and it is aligned with what we proved in Section \ref{sec:SDT}.
The more surprising insight from the experiment is that
introducing lower bounds that look just one or two steps ahead (i.e., $S_1$ and $S_2$) makes a great improvement in terms of performance of the $A^*$ approach, without affecting the computation time significantly.

Second, it can be seen that the box is concentrated around 0 for each lower bound, which implies that most instances are solved within a few hundred nodes. What influences the
average number of nodes are the ``hard-to-solve'' cases where a large number of nodes is
needed.
Using $S_{21}$ instead of $S_0$ makes the $A^*$ algorithm more efficient by
decreasing the number of nodes by a factor of 3 for those cases. From now on, we use $S_{21}$ as the lower bound.
Recall that actually we only need to compute $S_k$ for $k \le N-C$, as $S_{k+1}, \dots, S_N$ are all equal to $S_{k}$. Taking advantage of
this property further improves computational tractability.

\paragraph{The convergence rate of the lower bound.}
After introducing the lower bounds, we have shown in Propositions~\ref{prop:path1} and~\ref{prop:path2} that
on any given path, the lower bound is non-decreasing. The efficiency of the algorithm mainly
depends on the rate of increase of the lower bound on each path. If this rate is
high, then the algorithm has a greater chance to find the optimal solution with fewer nodes, as the tree is pruned faster. The lower bounds can either ``collapse'' with the upper bounds, or exceeds the best incumbent available.

\begin{figure}[h]
\centering
\includegraphics[width=0.45 \textwidth]{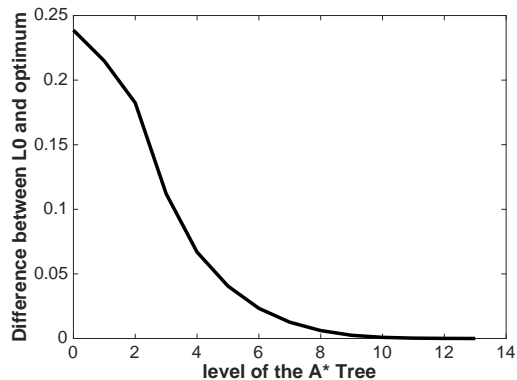}
\caption{The convergence rate of the lower bound on the optimal path}
\label{fig:Exp2}
\end{figure}

In this experiment and for the sake of clarity, we only focus on the rate of convergence of the lower bound
on the optimal path. For each instance, we find the optimal path, and at each
level $l$, we compute the difference between $S_{21}(B^l)$ and $z_{opt}(B^l)$. We
average this difference on all instances and plot the results in Figure~\ref{fig:Exp2}

First of all, we can see that even at level 0 (meaning at the root node), the
average difference is less than 0.25. This shows that on average the lower bound
is close to optimal, therefore it does not need to increase a lot to reach the
optimal solution. Second, we can see that after level 10, the average gap is less than 0.001, meaning that
after 10 relocations, in most cases, the
lower bound has reached optimality on the optimal path. Therefore, this
path will terminate at level 10 if the upper bound also collapsed to the
optimal. Finally, the trend of the curve shows that the gap between optimal and
the lower bound decreases faster at the beginning of the tree.
A simple way to explain this is that the lower bound increases when a future ``bad'' move can be predicted. But such bad moves become harder to predict as the bay gets emptier (there are more empty columns in the bay).

\begin{figure}[h]
\centering
\subfloat[Average Gap as a function of $\mathcal{N}$]
{\includegraphics[width=0.4 \textwidth]{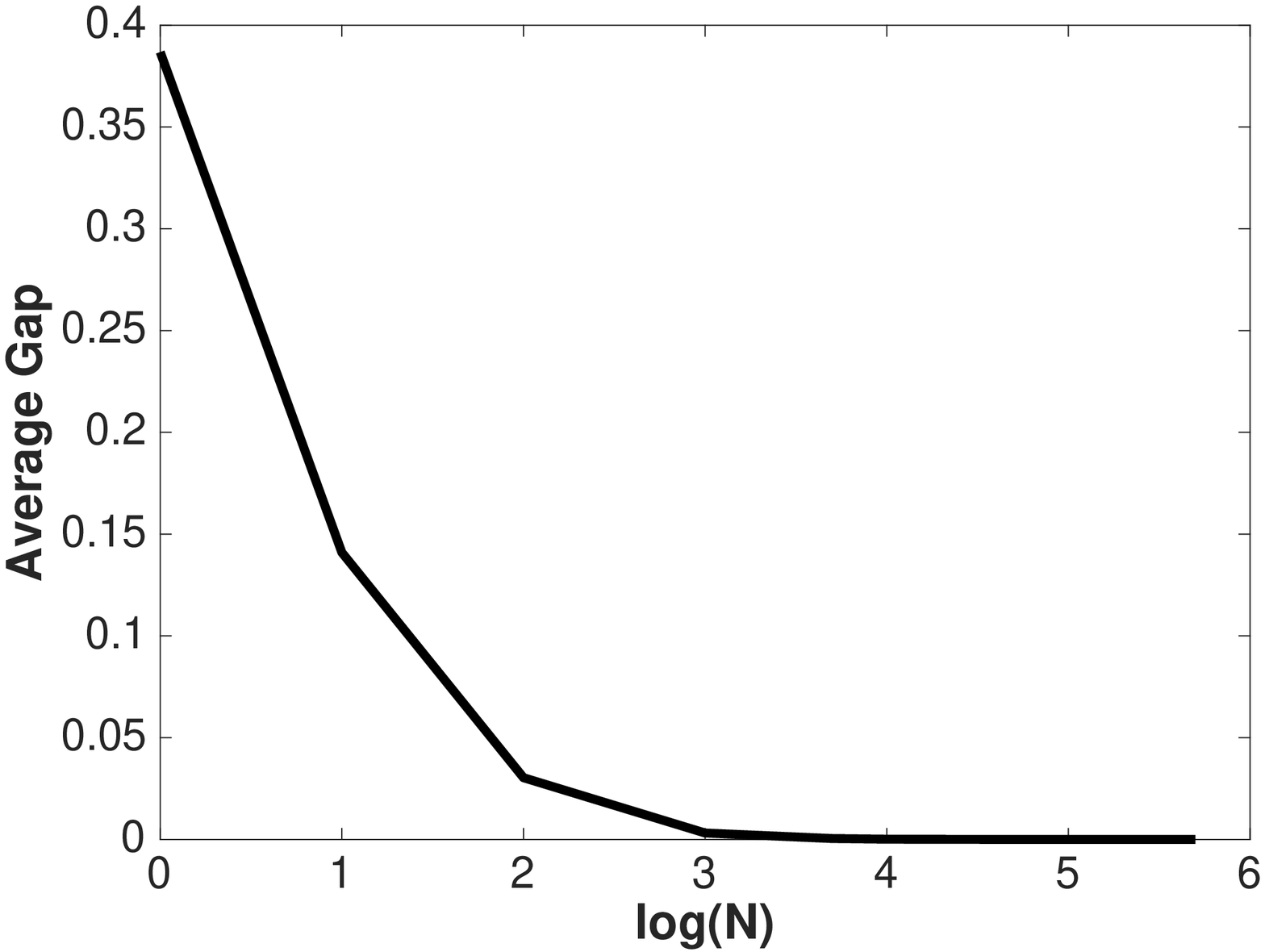}}
\quad \hspace{1cm}
\subfloat[Percentage of instances solved optimally as a function of $\mathcal{N}$]
{\includegraphics[width=0.4 \textwidth]{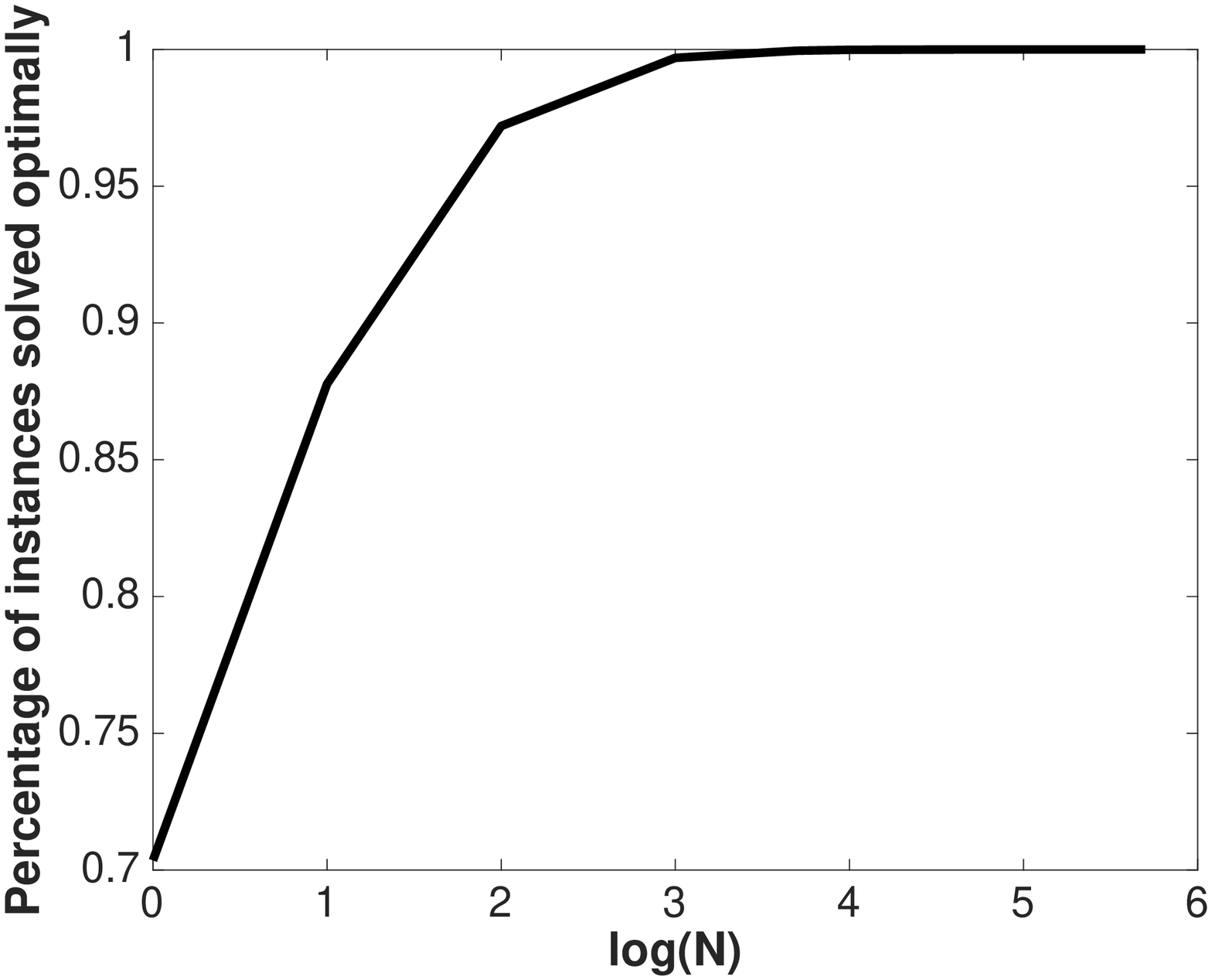}}
\caption{The Effect of $\mathcal{N}$}
\label{fig:Exp3}
\end{figure}

\paragraph{The effect of $\mathcal{N}$.}
By construction, the $A^*$ method is tunable. In this experiment, we study the impact of $\mathcal{N}$ on the average guaranteed gap with optimality,
and the percentage of instances solved optimally.
Notice that from Propositions~\ref{prop:path1} and~\ref{prop:path2}, we know that
the guaranteed gap is a non-increasing function of $\mathcal{N}$.
We solve each instance with different $\mathcal{N}$ and we record the gap with the
optimal solution obtained by the $A^*$ algorithm. We show the results in Figure~\ref{fig:Exp3}. On the left, we give
the average gap as a function of $\mathcal{N}$ and on the right, we show the percent of instances solved optimally.

First, we observe that the average guaranteed gap at the root node is very small (less than 0.4). Also,
in more than 70\% of instances, the optimal solution is found at the root node.
Second, as we increase $\mathcal{N}$, the gap sharply decreases and the number of instances solved to optimality increases.

This experiment is very useful from a practical a point of view in order to determine $\mathcal{N}$.
We can set an average gap that we allow in our solutions and infer the minimum $\mathcal{N}$
that is needed to achieve this performance on average. For example for bays of 4 by 7, if we allow
for a gap of 0.05 on average, we can set our $\mathcal{N}$ to 100.

This experiment is related to the first experiment. As shown in
Figure~\ref{fig:Exp1}, most instances only require few hundreds of
nodes, but there are some that require hundreds of thousands. This explains the
``concavity'' of the function in the left plot and the ``convexity'' of the function in the right plot.
In conclusion, this experiment highlights the trade-off between the quality of the solution (represented by the average guaranteed gap)
and its tractability (represented by $\mathcal{N}$).

\paragraph{Benchmark of a new heuristic and existing heuristics.}
Many heuristics have been developed for the C\@R\@P. From a practical point of view, heuristics are valuable since they are fast and easy to implement in container terminals. Therefore, it
is relevant to evaluate their performance to advise operators.
In order to benchmark existing heuristics, one can measure:
\begin{itemize}
\item The distribution of the difference between heuristics and the optimal solution; and
\item The expected performance ratio of the heuristic defined as
\begin{eqnarray}
  PR(heuristic)=\frac{z_{heuristic}-z_{opt}}{z_{opt}},
\end{eqnarray}
where $z_{heuristic}$ is the number of relocations in the heuristic solution. PR shows the relative gap
with optimality for a given heuristic.
\end{itemize}

From Figure \ref{fig:Exp1}, if we set $\mathcal{N}$ to $400,000$, then all $100,000$ instances are solved optimally using the $A^*$ algorithm. In addition to
heuristic H, we study two other existing heuristics presented by Kim and Hong (KH from~\cite{ref5} which uses the
estimation of future relocation) and Petering and Hussein (LA-5
from~\cite{ref9} that takes into account ``repositioning moves'', i.e. repositioning containers that are not in the same column as the target container).

\begin{table}[ht]
  \centering
\ra{1.2}
\begin{tabular}{@{}cccccccccccccccc@{}}\toprule
  \textit{Gap with} && $\textit{Heuristic H}$ && $\textit{Heuristic KH}$ && $\textit{Heuristic LA-5}$ && $\textit{Heuristic TH-2}$\\
 \textit{Optimal} && \textit{Distribution} && \textit{Distribution} &&  \textit{Distribution}&&
 \textit{Distribution}\\
\midrule
0 && 87.0\%  && 31.8\% && 83.5\% && \textbf{95.7\%}  \\
1 && 11.4\% && 21.4\% && 13.7\% && \textbf{3.98\%}\\
2 && 1.4\%  && 21.3\% && 2.0\% && \textbf{0.3\%} \\
$\ge\ 3$ && 0.2\% && 11.4\% && 0.3\% &&  \textbf{0.02\%}\\  \midrule\
$\mathbb{E} \left[{ \textit{PR}}\right]$ && 1.44\% && 16.0\% && 1.81\%
&& \textbf{0.44\%} \\
\bottomrule
\end{tabular}
\caption{Benchmarks of Heuristics on 100,000 instances}
\label{ref:bench}
\end{table}

We use the same idea as in the $A^*$ algorithm and introduce a new class of heuristics that can improve any existing heuristics; we refer to this class as Tree Heuristic (TH-L).
The basic idea of TH-L is to take the L best columns and branch on them to construct a decision tree. The L best columns can be chosen using any of the existing heuristics that compute a score for each column. Using this principle and considering several good candidates for each relocation, we are less likely to make a mistake. Here, we implement the TH-L with the H heuristic. The algorithm is presented in
Algorithm~\ref{algo2}. In our experiment, we set $L=2$.
Note that T\@H results in less or the same number of relocations compared to H since the path of H
is included in the tree of TH-L.

\begin{algorithm}[ht]
\caption{Tree Heuristic}\label{algo2}
\begin{algorithmic}[1]
\Procedure{$[Z_{TH}]=TreeHeuristic(B,L)$}{}
\State\ $Z_{TH} \gets 0$
\While{B is not empty}
\BState\ \emph{Retrieval}:
\If{target container n is on top of its column}
\State\ \text{Retrieve n form B}
\BState\ \emph{Relocation}:
\Else\
\State\ \text{r $\gets$ topmost blocking container}
\State\ $C_1 \gets \underset{c_i \in \{ 1,\ldots,C \} \setminus c}{\text{sort arg increasingly}} \left\{ \min(c_i) | \min(c_i)>r \right\}$
\State\ $C_2 \gets \underset{c_i \in \{ 1,\ldots,C \} \setminus c}{\text{sort arg decreasingly}} \left\{ \min(c_i) | \min(c_i)<r \right\}$
\State\ $C_3 \gets [C_1,C_2]$
\State\ $S \gets C_3[1:L]$
\State\ $Z_{TH}=Z_{TH}+\underset{s \in S}{\min}\{ \text{TreeHeuristic (B where r moves to column s,L)} \}$
\EndIf\ 
\EndWhile\ 
\EndProcedure\
\end{algorithmic}
\end{algorithm}

Results for the four heuristics are summarized in
Table~\ref{ref:bench}. First, H is optimal
in most instances (87\%). This is one of the main reasons that heuristic H was chosen as an upper
bound in the $A^*$ method. Second, TH-2 is indeed closer to optimality
than H, KH and LA-5, in distribution and in terms of the average performance ratio.

Finally, notice that the percentages for LA-5 do not sum to
100\% since there are cases for which LA-5 is better than $A^*$, which is because
LA-5 considers ``repositioning moves'', i.e., it solves for the unrestricted CRP. As a result, for 0.5\% of the 100,000
instances, LA-5 solves the instance with one relocation less than $A^*$.

In our last experiment for complete information, we study how the parameter L affects the performance
of the TH-L heuristic.
We consider the same 100,000 instances and solve them with L varying from 1 to
6, and we record their performance ratio.
Notice that TH with L=6 considers all possibilities for the blocking container therefore it gives the same solution
as the $A^*$ without the use of bounds. Thus $PR(\text{TH-6})=0$. Further note that $L=1$ gives heuristic H.
\begin{table*}[h]
\centering
\ra{1.2}
\begin{tabular}{@{}lcccccccccccccc@{}}\toprule
L && 1 && 2 && 3  && 4 && 5 && 6\\
\midrule
PR (TH-L) && 1.44\% && 0.44\% && 0.27\%  && 0.20\% && 0.16\% && 0\%  \\ \bottomrule
\end{tabular}
\caption{Effect of parameter L on the performance of heuristic TH-L}
\label{ref:thheuristic}
\end{table*}

The main observation is that the marginal gain of branching is maximum when we increase L from 1 to 2.
By considering two promising columns instead of one, TH-2 finds better solutions for
most instances where H was not optimal. Note that increasing L (considering more candidate columns) will further improve the
solution; however the gain from more branching is small considering the exponentially increasing cost of computation.

\subsection{Experimental results for the CRP with Incomplete Information}
\label{sec 5-1}

\paragraph{Number of nodes and computation time for $\textit{\textbf{ASA}}^*$.} Stochastic optimization methods that are based on enumerating scenarios are usually computationally expensive. The $ASA^*$ algorithm, however, is fast and tractable due to the use of sampling and pruning that allow for suppressing many nodes in the tree. More importantly, the $ASA^*$ is tunable in the sense that one can set $\epsilon$ and $\delta$ to change the number of nodes and thereby solve an instance within a desired amount of time, and yet ensure that the loss from using large $\epsilon$ or small $\delta$ is bounded (as shown in propositions \ref{pro1}-\ref{pro3}).

\begin{figure}[htb!]
    \centering
    \subfloat[$ASA^*$ implemented with $\delta = 0.5$ and $\epsilon = 0.05$]{
        \includegraphics[scale=0.52]{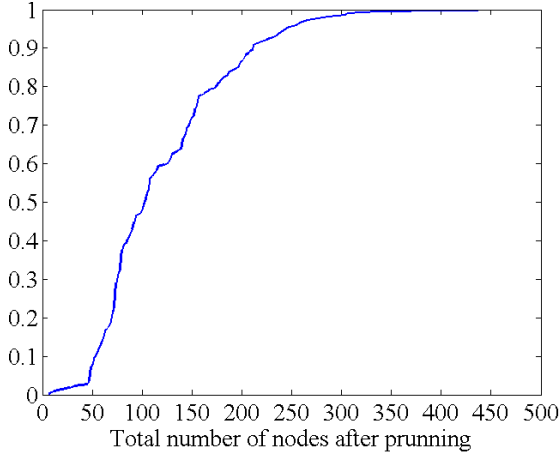}
\label{nodes}}
\hspace{2mm}
    \subfloat[Experiments were done on desktop computer with Intel i73770s processor (3.90 GHz) and 16Gb RAM.]{
        \includegraphics[scale=0.48]{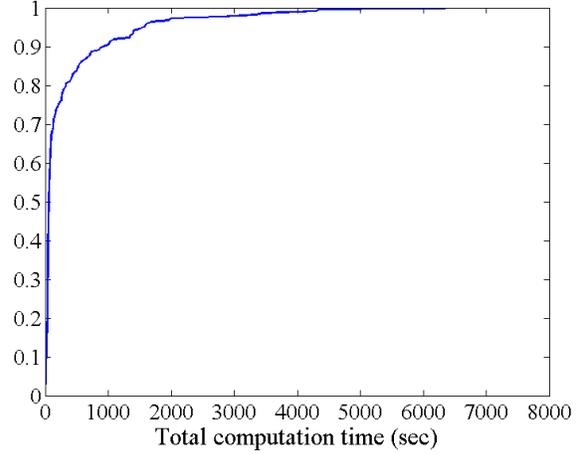}
\label{time}}
\caption{Tractability of the $ASA^*$ algorithm}
\label{computation}
\end{figure}

Figure \ref{nodes} shows the cumulative distribution of the number of nodes (after pruning) for the tree up to $t^*$=11, for a bay with 7 columns, 4 tiers, 3 containers per column, and $|\mathbf{K}|=11$ (Recall that $t^*$ is the time step at which the set of unknown containers become known and $|\mathbf{K}|$ is the number of containers that are initially known). We implement $ASA^*$ with $\delta = 0.5$ and $\epsilon = 0.05$. It can be seen that for half of the instances, the tree up to $t^*$ has 100 or fewer nodes. Also, about 90$\%$ of instances have 200 or fewer nodes. Note that the nodes of the trees that are constructed after time $t^*$ for solving instances corresponding to different scenarios do not have too much of an effect on computation time because the $A^*$ algorithm is very fast.
Figure \ref{time} shows the cumulative density of the average computation time. It can be seen that half of the instances are solved in less than two minutes and 90$\%$ of the instances are solved in 15 minutes or less.

Although the $ASA^*$ algorithm is fast for medium-sized bays, it might not be tractable for large bays. We next introduce a myopic heuristic and use both $ASA^*$ and the heuristic to provide several insights on the value of information.

\textbf{Myopic Heuristic.}
This is an extension of the heuristic H that we explained in Section \ref{sec:SDT}. Similar to the setting that we explained in Section \ref{sec:intro}, the problem is to retrieve $N$ containers from a bay with $C$ column and $P$ tiers. In the incomplete information setting, the departure order of containers $\{1, 2, \dots, |\mathbf{K}|\}$ are known at time zero; the departure order of containers $\{|\mathbf{K}|, |\mathbf{K}|+1, \dots,N\}$ become known at time step $t^{*}>0$.

Suppose $n$ is the target container located in column $c$. If $n$ is not blocked by any container, it is retrieved without any relocation. Otherwise, let $r$ be the topmost blocking container in column $c$, and $c^*$ be the column where r should be relocated. The myopic heuristic determines $c^*$ using the same rules as the H heuristic explained in Section \ref{sec:SDT}, except that we assign an index of $N+1$ to all unknown containers and we set $min(c_i)=N+2$ if $c_i$ is empty. Ties are broken arbitrarily. In the following experiments we study myopic heuristic and compare its performance to that of ASA*.

\paragraph{Value of information.}
To study the effect of the level of information initially available ($|\mathbf{K}|$), we take 1000 random instances of a bay with 7 columns, 4 tiers, $N=21$ containers, and 3 containers per column. We solve each instance with 6 levels of information: $|\mathbf{K}|$ = $\lceil$0.25N$\rceil$, $\lceil$0.375N$\rceil$, $\lceil$0.5N$\rceil$, $\lceil$0.625N$\rceil$, $\lceil$0.75N$\rceil$, and $\lceil$0.9N$\rceil$, using the $ASA^*$ algorithm and the myopic heuristic ($\lceil x \rceil$ is the smallest integer larger than $x$). For all cases, we fix $t^*$ at $\lceil$0.25N$\rceil+1$ to ensure that at every time-step before $t^*$, at least one container is known. For each of the 12 cases, we compute the average of relocations over the 1000 instances. We then compare the relative gap of each case with the average relocation for the CRP with full information: $\frac{\mathbb{E}[z_{ASA^*}]-\mathbb{E}[z_{opt}]}{\mathbb{E}[z_{opt}]}$ and $\frac{\mathbb{E}[z_{MH}]-\mathbb{E}[z_{opt}]}{\mathbb{E}[z_{opt}]}$, where $z_{MH}$ is the number of relocations obtained by the myopic heuristic.

\begin{figure}[htb!]
    \centering
    \subfloat[Value of information: relative gap in relocations \textcolor{white}{text}     with the case of full information]{
        \includegraphics[scale=0.44]{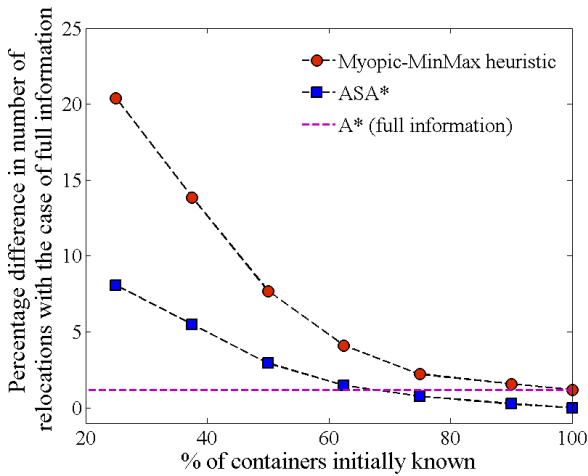}
\label{gapInfo}}
\hspace{1mm}
    \subfloat[Comparing the myopic heuristic and the $ASA^*$ algorithm: relative gap in relocations for different levels of information]{
        \includegraphics[scale=0.45]{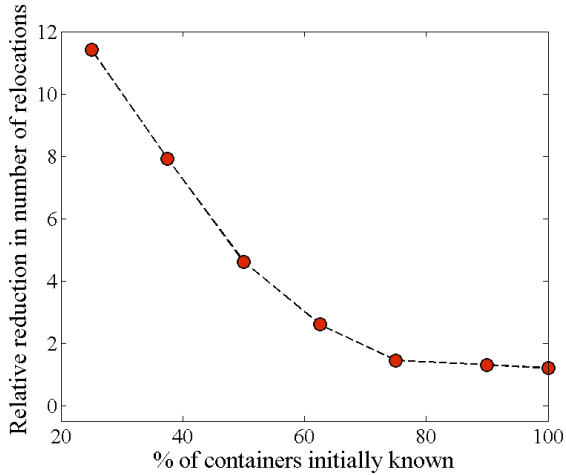}
\label{gapHeuDP}}
\caption{Experiment results for 7$\times$4 bay}
\label{gaps}
\end{figure}

Figure \ref{gapInfo} shows the relative gap in relocations for different levels of information. With the $ASA^*$ algorithm (implemented with $\delta$ = 0.5 and $\epsilon$ = 0.05), the gap is about 8$\%$ when 25$\%$ of the containers (6 containers) are known at time zero. The gap reduces to 3$\%$ when half of the containers are initially known and is almost zero when 90$\%$ of containers are known at time zero. The same behaviour can be observed for the heuristic. In both cases, the marginal value of information becomes smaller when more information is available. This is more significant for the heuristic. For example, when the level of information increases from 25$\%$ to 50$\%$, there is a significant drop in the gap; then the gap decreases more slowly and approaches zero at 100$\%$ information. Note that with 100$\%$ information, the myopic heuristic is the same as the H heuristic.

To get an insight into the value of information for the myopic heuristic, recall that it behaves similar to the H heuristic as long as $min(c_i)$ for all columns are known. Since all unknown containers have larger indices than the known containers, knowing at least one container in each column is sufficient to obtain the same $min(c_i)$ as in the heuristic H. Thus, after some point, having more information does not have much of effect on the number of relocations when we use the myopic heuristic.

Figure \ref{gapHeuDP} shows how the gap between the myopic heuristic and $ASA^*$ shrinks as the level of information increases. When 25$\%$ of the containers are initially known, using the myopic heuristic results in 12$\%$ more relocations on average (compared to $ASA^*$). This gap drops to less than 2$\%$ when all containers are known. Note that when more than 50\% of the containers are initially known, the solution provided by the myopic heuristic is reasonably close to the solution of $ASA^*$ (less than 5\%). Therefore, the myopic heuristic can be used in practice as it is easy to implement and efficient.

\begin{figure}[htb!]
\centering
{\includegraphics[width=10cm,height=7cm]{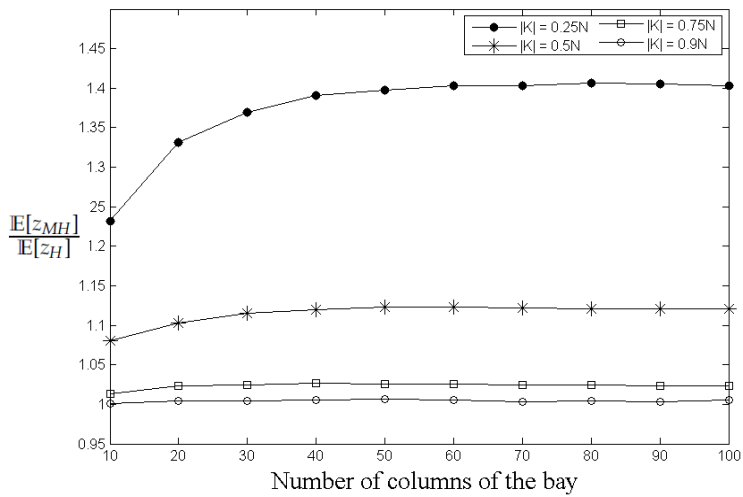}}
\caption{Asymptotic behaviour of myopic heuristic: $\frac{\mathbb{E}[z_{MH}]}{\mathbb{E}[z_H]}$ approaches a constant as the bay gets larger.}
\label{infoBays}
\end{figure}

It is interesting to see the effect of different levels of information on the number of relocations in larger bays. Figure \ref{infoBays} shows the average relative gap in relocations for bays of different sizes. For each bay size, the average relative gap is computed for 100,000 instances, using the myopic heuristic. The most important observation is that the ratio of $\mathbb{E}[z_{MH}]$ and $\mathbb{E}[z_H]$ does not constantly increase with bay size. And, for each information level, this ratio converges to a constant as the bay gets larger. Two other observations can be made from Figure \ref{infoBays}. First, the converging ratio for large bays drops fast as the level of information increases. For example, when 25$\%$ of containers are initially known, the ratio converges to $1.4$. This number drops to $1.03$ when the level of information is 75$\%$. Second,
the rate of convergence is much faster when more information is available. For example, when 75$\%$ or more of the containers are initially known, the ratio is almost constant independent of the bay size. When 25$\%$ of the containers are known, the ratio converges to 1.4 for bays with 40 or more columns. Last, note that the asymptotic behaviour of the myopic heuristic in Figure \ref{infoBays}, is as an upper bound on the performance of the $ASA^*$ algorithm.

What is shown in Figure \ref{infoBays} provides useful insights for port operators. Considering the gain (fewer relocations) at each level of information, port operators can design appointment systems that capture this gain by offering faster or discounted service to the customers who provide their arrival time in advance.

\paragraph{Comparing the myopic heuristic with existing heuristic.} To the best of our knowledge, only a few recent papers studied CRP with incomplete information, and developed heuristic algorithms for this problem \cite{dusan,RePEc:eee:transe:v:46:y:2010:i:3:p:327-343}. 
Here we compare our myopic heuristic with the RDH (revised difference heuristic of \cite{RePEc:eee:transe:v:46:y:2010:i:3:p:327-343}) and show that myopic outperforms the RDH. Zhao et al. \cite{RePEc:eee:transe:v:46:y:2010:i:3:p:327-343} show the percentage savings of RDH over a nearest relocation strategy for different bay sizes and for different amounts of information initially available. The greatest savings is about 50$\%$ and is realized for a bay with 12 columns and 7 maximum height, with 50 or more containers initially known (see Figure 13 of \cite{RePEc:eee:transe:v:46:y:2010:i:3:p:327-343}). The total number of containers is not indicated for this simulation, so we assume the bay is full with 72 containers. (With 50 known containers, the level of information is about 70$\%$). For a bay with 12 columns and 3 containers per column, the maximum savings is about 40$\%$ and is realized when 22 or more containers are known (i.e., when the information level is about 60$\%$).

\begin{figure}[htb!]
\centering
{\includegraphics[width=10cm,height=7cm]{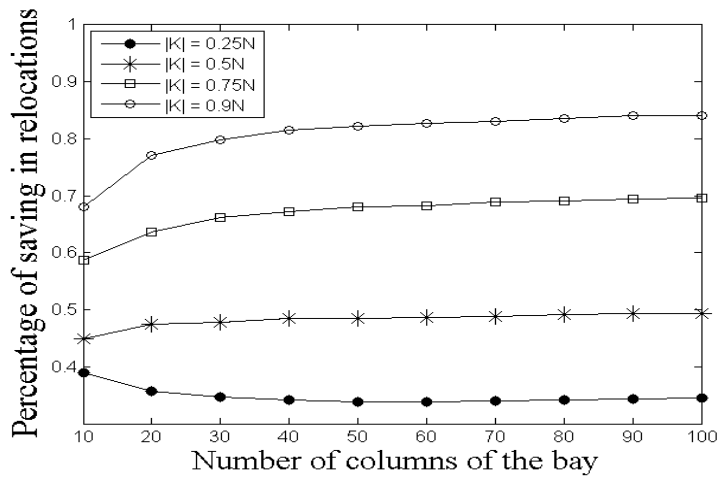}}
\caption{Percentage of saving of myopic heuristic over nearest relocation strategy.}
\label{rdh}
\end{figure}

In Figure \ref{rdh}, we show the percentage savings of the myopic heuristic over the nearest location strategy for bays with different numbers of columns, with 4 tiers and 3 containers per column. The savings are shown for 4 different levels of information. It can be seen that when level of information is 50$\%$, the saving is about 48$\%$, which is greater than the saving for RDH when even more information (about 70$\%$) is available. When the level of information is 25$\%$, the savings of the myopic heuristic is about 35$\%$, whereas for the same setting, RDH results in less than 25$\%$ savings over the nearest relocation strategy.

\section{Conclusion}
\label{disc}

Managing the relocation moves is one of the main challenges in the storage yard of container terminals and has a direct effect on the costs and efficiency of yard operations. The Container Relocation Problem (CRP) is notorious for its computational intractability and most research studies have designed heuristics in order to solve the problem, particularly for large bays.

In this paper, we revisited a fast and efficient method, the $A^*$ algorithm that provides optimal solution for the CRP. We proved several properties of the method as applied to the CRP and assessed its performance and computational tractability through experiments. We also studied the asymptotic behavior of optimal solutions of the CRP and showed that the ratio of the optimal number of relocations and a simple counting lower bound converges to 1 as the bay gets very large. This gives strong evidence that CRP is easier in large bays. Moreover, we showed through experiments that the H heuristic (introduced in \cite{RePEc}) has the same behavior. An important result of our experiments is that it is recommended to apply the heuristic H to large blocks of containers (several bays) in order to minimize the loss from using a heuristic rather than optimization methods (that are computationally expensive). We also presented an improvement over H heuristic, and show through experiments that it outperforms existing heuristics.

Further, we extended our study to the CRP with incomplete information, which is an important case in practice because the retrieval order of containers are usually not known far in advance. We introduced a two-stage approximate stochastic optimization framework ($ASA^*$)
that optimizes the expected number of relocations given a probabilistic distribution on the departure order of containers. We also presented a myopic heuristic that is very fast and efficient for the CRP with incomplete information. We used $ASA^*$ and the heuristic to study the value of information for different bay sizes and showed that the relative gap between $ASA^*$ and the heuristic shrinks as the level of information increases. In fact, when the available information is more than 50\%, the myopic heuristic is very efficient and can be used in practice.
We also showed that the ratio of the number of relocations between heuristic H and the myopic heuristic converges to a constant fast as bays get larger, which again implies that the loss from using the myopic heuristic is minimized when applied to large bays.

Our paper opens directions to many further studies in both complete information setting and the incomplete one. A very interesting problem in the complete information case is to generalize our model to a setting where stacking and retrieving processes overlap in time. In such a setting, the goal would be to jointly optimize the number of relocation for stacking and retrieving. We believe that one of the major questions of this problem is how to prioritize between stacking and retrieving at a given time.
In the incomplete information case, one can easily generalize our two-stage stochastic optimization problem to a
multi-stage setting where ASA* could be extended. However, as we increase the number of stages, the computation time grows rapidly. Additional assumptions and/or other efficient heuristics would be needed to ensure a reasonable running time.
Finally, another interesting problem is to study the case where no information about any containers is known in advance. Intuitively, it appears that the policy of placing the container on the emptiest column should be the optimal solution; however this remains to be proven.

Other storage systems such as steel plate stacking and warehousing systems face the relocation problem with complete and incomplete information (see Kim et al. \cite{Lastparagraphref1}, Z{\"a}pfel and Wasner \cite{Lastparagraphref2} for the former, and Chen et al. \cite{Lastparagraphref3} for the latter). We believe that the frameworks proposed in this paper could also be applicable to those problems.


\bibliographystyle{apa}
\bibliography{refs}
\newpage
\section*{Appendix}

\begin{proof}[\textbf{Proof of Proposition \ref{prop:path1}}]
  Between level $l$ and $l+1$, one move has been operated ($r$ moved to column $c_i$).
  If $r<\min({c_i})$ (`good'' move) then $r$ was a blocking container in $B^l$ but not in $B^{l+1}$, thus $S_0(B^{l+1})=S_0(B^l)-1$.
  If $r>\min({c_i})$, we have $S_0(B^{l+1})=S_0(B^l)$.
  So $S_0(B^{l+1})=S_0(B^l)-\chi(r<\min({c_i}))$ where $\chi$ is the indicator function.
  Therefore
  $L_0(B^{l+1})= S_0(B^{l+1})+(l+1)=S_0(B^l)-\chi(r<\min({c_i}))+1+l \ge S_0(B^l) +l = L_0(B^{l})$.
\end{proof}

\begin{proof}[\textbf{Proof of Proposition \ref{prop:LB decreasing}}]
  We have
  $L_p(B)-L_{p-1}(B) = \sum_{r \in \mathcal{R}_p(B)} \chi(r>MM(B_p))$. Since $\chi$ is a non negative
  function, the results follows.
\end{proof}

\begin{proof}[\textbf{Proof of Proposition \ref{prop:augmentingpathLN}}]
 Intuitively, this property comes from the fact that $S_{p} (B^{l+1}) \ge
 S_p(B^{l}) -1 $. Indeed, if this holds, we have
 $$L_p (B^{l+1})=S_p (B^{l+1})+l+1 \ge S_p(B^l)-1+l+1=S_p(B)+l=L_p(B)$$
 So let us prove $S_{p} (B^{l+1}) \ge  S_p(B^{l}) -1$. Let $r$ be the container
 that is relocated between $B^l$ and $B^{l+1}$. We have two cases:
 \begin{itemize}
   \item If $r$ can do a good move, $r$ has only been counted once in $S_p(B^l)$ as it is a
   blocking container in $B^l$ and it might or might not contribute to $S_p(B^{l+1})$.
   Therefore $S_p(B^{l+1}) \ge S_p(B^{l})-1$.
   \item  If $r$ cannot do a good move, it has been counted twice in $S_p(B^l)$.
   But since it cannot do a good move, it is necessarily a blocking container in
   $B^{l+1}$ so it counts at least once in $S_p(B^{l+1})$. Thus $S_p(B^{l+1}) \ge
   S_p(B^{l})-1$.
 \end{itemize}
 This completes the proof.
 \end{proof}

 \begin{proof}[\textbf{Proof of Proposition \ref{prop:caseCcontainers}}]
  Using heuristic H, a container is relocated only if it is
  blocking. For any blocking container, there exists
   at least one empty column which means there is a possible good move. So each blocking container
   is relocated at most once, i.e, $z_{H}(B) = S_0(B)$. Since $S_0(B) \le z_{opt}(B) \le z_{H}(B) = S_0(B)$
  and we have equation (\ref{caseCcontainers}).
\end{proof}

\begin{proof}[\textbf{Proof of Proposition \ref{prop:caseC1containers}}]
For simplicity of proofs we assume that $n_1$, the target container in the bay is 1.
  If 1 is not blocked then it is retrieved. In that case, there are C containers left
 and thus we use Proposition~\ref{prop:caseCcontainers} (note that in that case
 $S_1(B)=S_0(B)$).

  Then let us suppose 1 is blocked by some containers. If there is more than
  one blocking container, all except the last one can be relocated to empty columns.
  So we come back to the case where 1 is only blocked by
  container r, and there is no empty column.

 Now we consider two cases: If $r$ can do a good move and H relocates $r$ to a good
 column and then retrieves 1. We call $B'$ the new configuration. By
 Proposition~\ref{prop:caseCcontainers}, we have $z_H(B')=S_0(B')=S_1(B')$.
 Notice that $z_H(B)=z_H(B')+1$ and $S_1(B)=S_1(B')+1$ and the result follows.

 Now suppose $r$ cannot do a good move. This can only happen if $r=C+1$ and all the other
 columns have exactly one container. It is easy to check that in that case,
 $z_H(B)=S_1(B)$.
\end{proof}

\begin{proof}[\textbf{Proof of Proposition \ref{prop:caseCkcontainers}}]

First let us introduce some notations. Let $B$ be the initial configuration and $B'$ be the configuration obtained after retrieving container 1 using heuristic H. We denote $R_1$ the set of containers blocking container 1 in $B$.

  Now let us consider the case $k=2$. Note that using Proposition \ref{prop:caseC1containers} we have $z_H(B')=S_1(B')$ and $S_1(B') \le S_0(B') + 1$. There are four possibilities to consider:
   \begin{itemize}
  \item If $|R_1| = 0$, then it is easy to see that $z_H(B) = z_H(B') = z_{opt}(B') = z_{opt}(B)$.
  \item If $|R_1| = 1$, we know that $S_0(B') \le S_0(B)$. Therefore $z_H(B) = 1 + z_H(B') = 1 + S_1(B') \le 1 + S_0(B') + 1 \le S_0(B) + 2 \le z_{opt}(B) + 2$.
  \item If $|R_1| > 2$, then there are at least $|R_1|-2$ columns that are empty in B. Therefore the $|R_1|-2$ topmost containers blocking 1 find necessarily a "good" column in B. Hence this case can be reduced to the case where $|R_1|=2$.
  \item If $R_1=2$, then we claim that $S_1(B') \le S_0(B)$. If this is true, then $z_H(B) = z_H(B') + 2 = S_1(B') + 2 \le S_0(B) + 2 \le z_{opt}(B) + 2$. Now let us denote $r_1$ and $r_2$ the two containers blocking 1 in $B$.
  \begin{itemize}
  \item If $r_1 \le C$ or $r_2 \le C$ then $S_0(B') \le S_0(B) - 1$ and therefore $S_1(B') \le S_0(B')+1 \le S_0(B)$.
  \item If $r_1 > C$ and $r_2 > C$, then if one of them finds an empty column then again $S_0(B') \le S_0(B) - 1$. Otherwise it means that all columns have one container and $C+1$ and $C+2$ are on top of 1. In those two bays we can verify that $ z_H(B) = 4 \le 2+ 2 = S_0(B)+2 \le z_{opt} + 2$.
  \end{itemize}
  \end{itemize}
  This concludes the case $k=2$.

   For $3 \le k \le C$, the proof is simpler and works by induction.
  For $k=2$, the previous paragraph proves a stronger inequality. Let us suppose it is true for $k-1$,
  and $B$ has $C+k$ containers.
  We have by induction hypothesis, $z_H(B') \le S_0(B')+\frac{k(k-1)}{2}$.
  We have three cases:
  \begin{itemize}
  \item If $|R_1| = 0$, then $z_H(B) = z_H(B') \le S_0(B')+\frac{k(k-1)}{2} \le S_0(B) + \frac{k(k+1)}{2} \le z_{opt}(B)+\frac{k(k+1)}{2}$.
  \item If $|R_1| > k$, then there are at least $(C-1)-(C+k-(|R_1|+1))=|R_1|-k$ empty columns and therefore $S_0(B') \le S_0(B)+k-|R_1|$ which implies $z_H(B)  =|R_1|+z_H(B') \le |R_1| + S_0(B')+\frac{k(k-1)}{2} \le |R_1| + S_0(B)+k-|R_1| + \frac{k(k-1)}{2} = S_0(B)+\frac{k(k+1)}{2} \le z_{opt}(B)+\frac{k(k+1)}{2}$.
  \item If $0 < |R_1| \le k$, then since $S_0(B') \le S_0(B)$ we have $z_H(B)  =|R_1|+z_H(B') \le |R_1| + S_0(B')+\frac{k(k-1)}{2} \le k + S_0(B) + \frac{k(k-1)}{2} = S_0(B)+\frac{k(k+1)}{2} \le z_{opt}(B)+\frac{k(k+1)}{2}$.
  \end{itemize}

which concludes the proof.

\end{proof}

 \begin{proof}[\textbf{Proof of Proposition \ref{lemmalinlowerbound}}]
   Let $S_0^i (B_C)$ be the number of blocking containers in column $i$. By the linearity of expectation, we have
   \begin{eqnarray*}
   \mathbb{E}_C\left[ S_0 (B_C)\right] = \mathbb{E}_C\left[ \sum_{i=1}^C S_0^i (B_C)\right] = \sum_{i=1}^C
   \mathbb{E}\left[ S_0^i (B_C)\right] = \alpha_h \times\ C,
   \end{eqnarray*}
    where $\alpha_h = \mathbb{E}\left[ S_0^1(B_C)\right]$. The last equality comes from the fact that each column is identically
   distributed.

   Now let us compute $\alpha_h$. By definition, $\alpha_h = \sum_{k=1}^{h-1} k p_{k,h}$.
   Let us start with
   $p_{0,h}$. Given h randomly chosen containers, the probability that they are
   placed in the first column in a descending order (meaning there is no blocking container)
   is $p_{0,h}=\frac{1}{h!}$. Now for $1 \le k \le h-1$, we compute $p_{k,h}$ by recursion (conditioning on the position of the smallest container). Let $n^*$ be the smallest container among these h containers. $n^*$ is located in $j^{th}$ topmost tier with probability $\frac{1}{h}$. Conditioned on that event, it means that $(j-1)$ containers are blocking $n^*$. So if $j>k+1$, then the event of having $k$ blocking container is not possible. Therefore for $j \le k+1$ there should be $(k-(j-1))$ blocking containers below the $j^{th}$ topmost tier and this happens with probability
   $p_{k-(j-1),h-j}$ (since there are $h-j$ tier below the $j^{th}$ topmost tier). Summing over all $j$ (possible positions of $n^*$), we have
   \begin{eqnarray*}
     p_{k,h}=\sum_{j=1}^k \frac{1}{h} p_{k-j+1,h-j},
     \end{eqnarray*}
     which completes the proof
 \end{proof}

 \begin{proof}[\textbf{Proof of Theorem~\ref{theor:asym}}]
  Since for all configurations $B_C$, $z_{opt} (B_C) \ge S_0(B_C)$ then
  $\frac{\mathbb{E}_C\left[ z_{opt} (B_C)\right]}{\mathbb{E}_C\left[ S_0 (B_C)\right]}
  \ge 1$.

   Moreover, we have:
    \begin{align}
     \frac{\mathbb{E}_C\left[ z_{opt} (B_C)\right]}{\mathbb{E}_C\left[ S_0 (B_C)\right]}
     & =1+ \frac{\mathbb{E}_C\left[ z_{opt} (B_C)\right] - \mathbb{E}_C\left[ S_0 (B_C)\right]}{\mathbb{E}_C
     \left[ S_0 (B_C)\right]}  \nonumber\\
     &= 1+ \frac{1}{\alpha_h C}\left( \mathbb{E}_C\left[ z_{opt} (B_C)\right] - \alpha_h C \right) \nonumber\\
     & = 1+\frac{g(C)}{\alpha_h C} \label{relationg}
   \end{align}
   where
   \begin{eqnarray}
     g(C)=\mathbb{E}_C\left[ z_{opt} (B_C)\right] - \alpha_h C.
   \end{eqnarray}
   In Lemma  \ref{fundlemma}, we study how $\mathbb{E}_C\left[ z_{opt} (B_C)\right]$
   evolves and we show that it increases almost linearly in $\alpha$ which shows
   that the function $g(.)$ is essentially bounded.
 \begin{lemma}
 \label{fundlemma}
     Let $z_{opt}(B_C)$ be the optimal number of relocations for configuration $B_C$,
     then there exists a constant $\theta$ (defined in equation (\ref{eqtheta})) such that:
     \begin{eqnarray}
       \label{fundlemmaeq}
        \mathbb{E}_{C+1}\left[ z_{opt} (B_{C+1})\right] \le \mathbb{E}_C\left[ z_{opt} (B_C)\right]
     + \alpha_h + h(P-1){(C+1)} e^{-\theta (C+1)} , \forall C \ge h+1
     \end{eqnarray}
   \end{lemma}
 Using Lemma~\ref{fundlemma}, we have for all $C \ge h+1$:
   \begin{align}
     & \mathbb{E}_{C+1}\left[ z_{opt} (B_{C+1})\right] \le \mathbb{E}_C\left[ z_{opt} (B_C)\right]
     + \alpha_h + h(P-1){(C+1)} e^{-\theta (C+1)}\nonumber\\
      \implies & \mathbb{E}_{C+1}\left[ z_{opt} (B_{C+1})\right] - \alpha_h (C+1) \le
     \mathbb{E}_C\left[ z_{opt} (B_C)\right] - \alpha_h C + h(P-1){(C+1)} e^{-\theta (C+1)}
     \nonumber\\
      \implies & g(C+1) \le g(C) + h(P-1){(C+1)} e^{-\theta (C+1)}
     \nonumber\\
      \implies & g(C) \le g(h+1) + h(P-1) \sum_{i=h+2}^{C} \left(i e^{-\theta i} \right)
      \le g(h+1) + h(P-1) \sum_{i=1}^{\infty} \left(i e^{-\theta i} \right)
     \nonumber\\
     \implies & g(C) \le g(h+1) + \frac{e^{\theta} h(P-1)}{{(e^{\theta}-1)}^2}=K'.
     \label{boundg}
   \end{align}
   Therefore using equations (\ref{relationg}) and (\ref{boundg}), we have
   \begin{eqnarray}
     \frac{\mathbb{E}_C\left[ z_{opt} (B_C)\right]}{\mathbb{E}_C\left[ S_0 (B_C)\right]}
     \le 1+ \frac{K}{C} = f(C),
   \end{eqnarray}
   where
   \begin{eqnarray}
   \label{constantK}
     K=\frac{K'}{\alpha} = \frac{g(h+1) + \frac{e^{\theta} h(P-1)}{{(e^{\theta}-1)}^2}}{\alpha}
   \end{eqnarray}
   which completes the proof of the theorem.
   \end{proof}

   \begin{proof}[\textbf{Proof of Lemma~\ref{fundlemma}}]
   Now we need to prove equation (\ref{fundlemmaeq}). Define a column to be ``special'' if all its containers
   are not smaller than $\omega=(h-1)(C+1)+1$.
   Now let us consider the following event:
 \begin{eqnarray}
   \label{event}
     \Omega =
     \left\{
     \begin{array}{l}{
     \textit{The bay with $C+1$ columns has at least one ``special'' column}}
     \end{array}
     \right\}.
 \end{eqnarray}

 The intuition is the following: the probability of having a ``special'' column
  grows quickly to 1 as a function of $C$ implying that the event $\Omega$ happens with high
  probability. Now, conditioned on $\Omega$, we
  more easily express the difference between bays of size $C+1$ and $C$ in the
  following way. We claim that
  \begin{eqnarray}
    \label{resultcondexp}
    \mathbb{E}_{C+1}\left[ z_{opt}(B_{C+1}) | \Omega \right] \le \mathbb{E}_C\left[ z_{opt}(B_C) \right]
    + \alpha_h,
  \end{eqnarray}

   Let $B_{C+1}$ be a given bay of size $C+1$ that verifies $\Omega$. Since columns in bays
     can be interchanged, we suppose that a ``special'' column is the first (leftmost) column of the bay.
     We also denote $n_1,n_2,\ldots,n_h$ the containers of
     the first column. We know that $n_1,n_2,\ldots,n_h \ge \omega$ and $n_1 \neq n_2 \neq \ldots \neq
     n_h$. Finally let $B_C$ be the bay $B_{C+1}$ without its first column (see
     Figure~\ref{fig:Baydecomp}).

     \begin{figure}[h]
     \centering
     \includegraphics[width=0.5\textwidth]{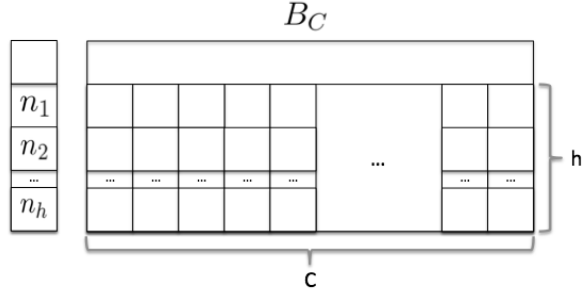}
     \caption{Bay decomposition of $B_{C+1}$ (The part on the right has $C$ columns)}
 \label{fig:Baydecomp}
     \end{figure}

     First we prove that
     \begin{eqnarray}
       \label{condituppbound}
       z_{opt}(B_{C+1}) \le z_{opt}(B_C) + S_0 \left(  \left[ \begin{array}{c}  n_1 \\ \ldots \\ n_h  \end{array}
      \right] \right).
     \end{eqnarray}
     To prove equation (\ref{condituppbound}), we construct a feasible
     sequence $\sigma$ for the bay of size $C+1$ for which the number of relocations is equal to the right side of
     equation (\ref{condituppbound}).
    Let $\sigma_{opt}(B_C)$ the optimal sequence for $B_C$, $t'$ be the first time step when
    the target container in $\sigma_{opt}(B_C)$ is larger than $\min\{ n_1,n_2,\ldots,n_h
    \}$ and $B'_C$ be the bay obtained at $t'$ using $\sigma_{opt}(B_C)$. Let the
    first $t'-1$ moves of $\sigma$ be the first $t'-1$ moves of
    $\sigma_{opt}(B_C)$.
    Note that $B'_C$ has at most $C-h$ containers due to the choice of $\omega$.
    By Proposition \ref{prop:caseCcontainers}, the number of
    relocations performed by $\sigma_{opt}(B_C)$ from $t'$ until the end is
    $S_0(B'_C)$. Therefore
    \begin{eqnarray}
      z_{opt}(B_C) = \textit{\# relocations up to $t'$ done by $z_{opt} (B_C)$} + S_0 (B'_C).
    \end{eqnarray}
    After $t'$, we run heuristic H on $B'_{C+1} = \left( \left[\begin{array}{c} n_1 \\ \ldots \\ n_h \end{array}
      \right] \cup\ B'_C \right)$.

      We claim that $z_{\sigma}$ is exactly the right side of equation
      (\ref{condituppbound}). There are at most $C$ containers in $B'_{C+1}$, therefore
      using Proposition \ref{prop:caseCcontainers}, we know that
      if we apply the heuristic H to this configuration, then the number of relocations done by
      H is $S_0(B'_{C+1})=S_0(B'_C)+S_0 \left(\left[\begin{array}{c} n_1 \\ \ldots \\ n_h \end{array}
      \right] \right)$. Therefore

    \begin{align*}
      z_{opt}(B_{C+1}) \le z_{\sigma}(B_{C+1})
  = \textit{\# relocations up to $t'$ done by $z_{opt} (B_C)$} + S_0(B'_C)+
      S_0 \left(\left[\begin{array}{c} n_1 \\ \ldots \\ n_h \end{array}\right] \right),
      \nonumber
    \end{align*}

    and we have
    \begin{eqnarray}
     z_{opt}(B_{C+1}) & \le z_{opt} (B_C) + S_0 \left(\left[\begin{array}{c} n_1 \\
         \ldots \\ n_h \end{array} \right] \right),
         \label{oneinstancebound}
    \end{eqnarray}

    which proves equation (\ref{condituppbound}).
    \\

    Now we can take the expectation from both sides of equation (\ref{resultcondexp}) over a uniform
    distribution of the rest of
    the $hC$ containers that are not in
    the first column. We claim that the first term on the right hand-side of equation (\ref{oneinstancebound})
    is exactly $\mathbb{E}_{C}\left[ z_{opt} (B)\right]$. For any configuration that appears
    in $B_C$, we can map it to a unique configuration where all containers are
    between 1 and $hC$. Thus,
    \begin{eqnarray}
     \label{Exprelation}
      \mathbb{E}_{C}\left[ z_{opt} (B_{C+1}) \left| \left[ \begin{array}{c} n_1 \\ \ldots \\ n_h  \end{array}
      \right. \right] \right]
      \le\ \mathbb{E}_{C} \left[ z_{opt} (B_C) \right]
      + S_0 \left( \left[ \begin{array}{c} n_1 \\ \ldots \\ n_h  \end{array}
      \right] \right).
    \end{eqnarray}

    Next, we take the expectation of both sides of equation (\ref{Exprelation})
    over possible first columns, which is a ``special'' column.
    Now notice that if $B_{C+1}$ is generated uniformly in the sets of bays of
    size $C+1$, then conditioned on $\Omega$, the probability of having a certain column
    ${\left[ n_1,\ldots,n_h \right]}^T$ is identical for any $n_1 \neq \ldots \neq n_h \ge \omega$
    and it is given by
    \begin{eqnarray}
      \mathbb{P}\left( \left. \left[\begin{array}{c} n_1 \\ \ldots \\ n_h  \end{array}\right] \right|  \Omega\ \right)
      =\frac{(C-h)!}{C!}.
    \end{eqnarray} Moreover, if we withdraw this
    first column from $B_{C+1}$ and look at the last $C$ columns (after renumbering the containers so that
    the numbers lie between 1 and $hC$), then this bay is
    also uniformly distributed in the sets of bays with $C$ columns.
    Therefore we can write:
    \begin{align}
      \mathbb{E}_{C+1}\left[ z_{opt}(B_{C+1}) | \Omega \right] & =
      \sum\limits_{\substack{(n_1,\ldots,n_h) \\ n_i \neq\ n_j \\ n_i \ge\ \omega}} \left(
      \mathbb{E}_{C}\left[ z_{opt} (B_{C+1}) \left| \left[\begin{array}{c} n_1 \\ \ldots \\ n_h  \end{array}
      \right] \right. , \Omega\ \textit{ }\right] \times\ \mathbb{P}\left(\left. \left[\begin{array}{c} n_1 \\ \ldots \\ n_h  \end{array}\right]
      \right|  \Omega\ \right) \right) \label{line1} \\
      & =
      \sum\limits_{\substack{(n_1,\ldots,n_h) \\ n_i \neq\ n_j \\ n_i \ge\ \omega}}
      \left(\mathbb{E}_{C}\left[ z_{opt} (B_{C+1}) \left| \left[\begin{array}{c} n_1 \\ \ldots \\ n_h  \end{array}
      \right] \right. \right] \times\ \mathbb{P}\left(\left. \left[\begin{array}{c} n_1 \\ \ldots \\ n_h  \end{array}\right]
      \right|  \Omega\ \right) \right) \label{line2} \\
      & \le\ \mathbb{E}_{C}\left[ z_{opt} (B_C)\right]
      \sum\limits_{\substack{(n_1,\ldots,n_h) \\ n_i \neq\ n_j \\ n_i \ge\ \omega}}
       \mathbb{P}\left(\left. \left[\begin{array}{c} n_1 \\ \ldots \\ n_h  \end{array}\right] \right|  \Omega\ \right)
      + \sum\limits_{\substack{(n_1,\ldots,n_h) \\ n_i \neq\ n_j \\ n_i \ge\ \omega}}
      S_0 \left(\left[\begin{array}{c} n_1 \\ \ldots \\ n_h  \end{array}
      \right] \right) \times\ \mathbb{P}\left(\left. \left[\begin{array}{c} n_1 \\ \ldots \\ n_h  \end{array}\right]
      \right|  \Omega\ \right)\nonumber\\
      & \le\ \mathbb{E}_{C}\left[ z_{opt} (B_C)\right]
      + \sum\limits_{\substack{(n_1,\ldots,n_h) \\ n_i \neq\ n_j \\ n_i \ge\ \omega}}
      S_0 \left(\left[\begin{array}{c} n_1 \\ \ldots \\ n_h  \end{array}
      \right] \right) \times\ \mathbb{P}\left(\left. \left[\begin{array}{c} n_1 \\ \ldots \\ n_h  \end{array}\right]
      \right|  \Omega\ \right)
 \label{eqalmostfinal}
   \end{align}
   The equality between \ref{line1} and \ref{line2} comes from the fact that if we know that $B_{C+1}$
   has a ``special'' column, then we do not need to condition on $\Omega$.
   Equation \ref{eqalmostfinal} uses the fact that $\sum\limits_{\underset{ n_i \ge\ \omega}{n_1 \neq\ \ldots \neq\ n_h}}
       \mathbb{P}\left( \left. \left[\begin{array}{c} n_1 \\ \ldots \\ n_h  \end{array}\right] \right|  \Omega\ \right)=1$.

    Note that, given any $ (n_1,\ldots,n_h)$ such that $n_i \neq\ n_j$, we have
    \begin{eqnarray*}
      \mathbb{E} \left[S_0 \left(\left[\begin{array}{c} n_1 \\ \ldots \\ n_h  \end{array}
      \right] \right) \right] = \alpha_h,
    \end{eqnarray*}
    when the expectation is over a random order of $ (n_1,\ldots,n_h)$. This is
    true regardless of the set $ (n_1,\ldots,n_h)$ that is drawn from
    (See Remark~\ref{remarkalpha}). This implies that the second term in the right hand side of
    equation (\ref{eqalmostfinal}) is equal to $\alpha_h$;
   Therefore, we get equation (\ref{resultcondexp}).

   Lemma~\ref{lemmaproba} states that the event $\Omega$ has a probability that increases
   exponentially fast to 1 as a function of C.

  \begin{lemma}
 \label{lemmaproba}
     Let $\Omega$ be the event defined by equation (\ref{event}), then there
     exists a constant $\theta>0$ such that
     \begin{eqnarray}
       \mathbb{P}(\overline{\Omega}) \le e^{-\theta (C+1)}
     \end{eqnarray}
     where $\theta$ is given by equation (\ref{eqtheta}).
   \end{lemma}

   Now we want to focus on the event $\overline{\Omega}$. We give an upper bound on
   $\mathbb{E}_{C+1} \left[ z_{opt}(B_{C+1}) | \overline{\Omega} \right]$. For any
   configuration, in order to retrieve one container, we need at most $P-1$
   relocations (since at most $P-1$ containers are blocking it), thus for any
   configuration, the optimal number of relocations is at most $P-1$ times the number
   of containers $(h(C+1))$ which gives us $h(P-1)(C+1)$ as an upper bound on the optimal number of
   relocations. We use this universal bound to get
   \begin{eqnarray}
     \mathbb{E}_{C+1} \left[ z_{opt}(B_{C+1}) | \overline{\Omega} \right] \le h(P-1)(C+1).
   \end{eqnarray}

  Finally using Lemma~\ref{lemmaproba}, we have
  \begin{align}
    \mathbb{E}_{C+1}\left[ z_{opt}(B_{C+1})\right]
    & = \mathbb{E}_{C+1}\left[ z_{opt}(B_{C+1}) | \Omega \right] \mathbb{P}(\Omega) +
    \mathbb{E}_{C+1}\left[ z_{opt}(B_{C+1}) | \overline{\Omega} \right]\mathbb{P}(\overline{\Omega})\nonumber\\
    & \le \mathbb{E}_{C+1}\left[ z_{opt}(B_{C+1}) | \Omega \right] +
    \mathbb{E}_{C+1}\left[ z_{opt}(B_{C+1}) | \overline{\Omega} \right] \times\ e^{-\theta (C+1)}\nonumber\\
    & \le \mathbb{E}_C\left[ z_{opt} (B_C)\right] + \alpha_h + h(P-1)(C+1)e^{-\theta (C+1)},
    \nonumber
  \end{align}
  which completes the proof of Lemma~\ref{fundlemma}.
  \end{proof}

  \begin{proof}[\textbf{Proof of Lemma~\ref{lemmaproba}}]
    Recall that
    \begin{eqnarray*}
     \Omega = \left\{
     \begin{array}{l}{
     \textit{The bay with $C+1$ columns has at least one ``special'' column}}
     \end{array}
     \right\}.
 \end{eqnarray*}

 We know that each bay of size $C+1$ can be mapped to a permutation $\pi$ of $\mathcal{S}_{h(C+1)}$
 taken uniformly at random. Let $q(.)$ be the
 function from $\mathcal{S}_{h(C+1)}$ to $\mathbb{R}^+$ defined by
 \begin{eqnarray*}
   q: \pi \longmapsto \textit{ number of ``special'' columns in the resulting bay of } \pi.
 \end{eqnarray*}
 Note that
 \begin{eqnarray*}
   \mathbb{P}\left( \overline{\Omega} \right) = \mathbb{P}\left( q(\pi)=0 \right).
 \end{eqnarray*}

 First we compute the expected value of $q(.)$
 \begin{align}
   \mathbb{E}_{C+1}[q] &=\mathbb{E}_{C+1} \left[\sum_{i=1}^{C+1} \chi \left(
   \textit{$c_i$ is a ``special'' column}\right)\right]\nonumber\\
    \mathbb{E}_{C+1}[q] &=(C+1) \times \mathbb{P} \left(\left\{ \textit{$c_1$ is a ``special'' column}
    \right\}\right),
 \end{align}
 where we use linearity of expectation and the fact that columns are identically distributed.

 A simple counting implies that
 \footnote{Notice that when $C \rightarrow \infty$, the probability is equivalent to $(\dfrac{1}{h})^h$ which would guarantee a faster convergence rate.}:
 \begin{align}
   \mathbb{P} \left(\left\{ \textit{$c_1$ is a ``special'' column}
    \right\}\right)&=\frac{(C+1)[(C+1)-1]\ldots[(C+1)-h+1]}{h(C+1)[h(C+1)-1]\ldots[h(C+1)-h+1]}
    \nonumber\\
    & \ge {\left(\frac{(C+1)-h+1}{h(C+1)}\right)}^h \nonumber\\
    & \ge {\left({\frac{2}{h(h+1)}}\right)}^h, \nonumber\\
 \end{align}
 where we use $C+1 \ge h+1$ to show the last inequality.
 \\

 Therefore we know that
 \begin{eqnarray}
   \label{expqfunction}
   \mathbb{E}_{C+1}[q] \ge (C+1) \times {\left({\frac{2}{h(h+1)}}\right)}^h.
 \end{eqnarray}

 We claim that $q(.)$ is well concentrated around its mean. In order to do so,
 we prove that $q(.)$ is 1-Lipschitz. Define $\rho$ the distance between
 two permutations $\pi_1,\pi_2 \in \mathcal{S}_{h(C+1)}$ as
 $\rho(\pi_1,\pi_2)=|\left\{ i \in [h(C+1)] : \pi_1(i) \neq \pi_2(i) \right\} |$.
 We want to prove that
 \begin{eqnarray}
    |q(\pi_1)-q(\pi_2)| \le \rho(\pi_1,\pi_2) , \forall \left(\pi_1,\pi_2\right) \in \mathcal{S}_{h(C+1)}.
 \end{eqnarray}
 Let $\pi_1,\pi_2 \in \mathcal{S}_{h(C+1)}$. Let us first consider the case where
 $\rho(\pi_1,\pi_2)=2$. (Notice that if $\rho(\pi_1,\pi_2) \neq 0$ then $\rho(\pi_1,\pi_2) \ge
 2$). In that case, we have $i,j \in \{1,\dots,n\}$ such that $\pi_1(i)=\pi_2(j)$ and
 $\pi_1(j)=\pi_2(i)$. Let $B_1$ and $B_2$ be the configurations generated by $\pi_1$ and $\pi_2$.
 Having $\rho(\pi_1,\pi_2)=2$ corresponds to the fact that if we swap 2 containers in $B_1$, we get $B_2$
 and we denote those containers $a=\pi_1(i)$ and $b=\pi_1(j)$. We have three cases:
 \begin{itemize}
   \item $a$ and $b$ are both in ``special'' columns in $B_1$. In this case, swapping
   them will not change anything since both their new columns in $B_2$ will also be
   ``special'' and hence $|q(\pi_1)-q(\pi_2)|=0$.
   \item $a$ and $b$ are both in columns that are not ``special'' columns in $B_1$. If $a,b \ge \omega$
   or $a,b < \omega$ then we do not create any new special column in $B_2$. Now
   suppose that $a \ge \omega$ and $b < \omega$, then the column of $a$ in $B_2$
   might be a ``special'' column, but the column of $b$ in $B_2$ cannot be ``special''.
   Therefore in that case, $|q(\pi_1)-q(\pi_2)| \le 1$.
   \item $a$ is in a ``special'' column in $B_1$ but $b$ is not.
   Now we know that $a \ge \omega$. If $b<\omega$ then the column of $b$ in $B_2$
   cannot be ``special'' but the column of $a$ might be and in that case $|q(\pi_1)-q(\pi_2)| \le 1$. If
   $b \ge \omega$, then the column of $b$ in $B_2$ is ``special'' and the column of $a$ in $B_2$
   is not ``special'' which gives us $|q(\pi_1)-q(\pi_2)| = 0$. Note that the proof
   is identical if $b$ is in a ``special'' column in $B_1$ but $a$ is not.
 \end{itemize}
 So far we have shown that
 \begin{eqnarray}
 \label{rho2}
 \textit{If } \rho(\pi_1,\pi_2)=2 \textit{, then } |q(\pi_1)-q(\pi_2)| \le 1.
 \end{eqnarray}

 Now we suppose that $\rho(\pi_1,\pi_2)=k$ where $2 \le k \le h(C+1)$. Note that
 we can construct a sequence of permutations $(\pi'_1,\pi'_2,\ldots,\pi'_k)$ such
 that $\pi'_1=\pi_1$, $\pi'_k=\pi_2$ and $\rho(\pi'_i,\pi'_{i+1})=2$.

 Now using this fact and equation (\ref{rho2}),
 \begin{eqnarray*}
   |q(\pi_1)-q(\pi_2)| = \left| \sum_{i=1}^{k-1} q(\pi'_i)-q(\pi'_{i+1}) \right|
   \le \sum_{i=1}^{k-1} |q(\pi'_i)-q(\pi'_{i+1})| \le \sum_{i=1}^{k-1} 1 = k-1 \le
   k = \rho(\pi_1,\pi_2),
 \end{eqnarray*}
 which proves that $q(.)$ is 1-Lipschitz.
 \\

 Now we use  Theorem 8.3.3 of Matou\v{s}ek and Vondr\'{a}k
 \cite{matouseklecturenotes} which states that
 $\mathbb{P}\left( q \le \mathbb{E}_{C+1}[q]-t \right)
   \le e^{- \frac{t^2}{8h(C+1)}}$
 and apply it with $t=\mathbb{E}_{C+1}[q]$ and equation (\ref{expqfunction}) to get
 \begin{eqnarray*}
   \mathbb{P}\left( q=0 \right) = \mathbb{P}\left( q \le \mathbb{E}_{C+1}[q]-\mathbb{E}_{C+1}[q] \right)
   \le e^{- \frac{{\left(\mathbb{E}_{C+1}[q]\right)}^2}{8h(C+1)}} \le e^{- \theta (C+1)},
 \end{eqnarray*}
 where
 \begin{eqnarray}
 \label{eqtheta}
   \theta = \frac{1}{8h} {\left(\frac{2}{h(h+1)}\right)}^{2h} > 0,
 \end{eqnarray}
 which concludes the proof.
  \end{proof}

 \begin{proof}[\textbf{Proof of Corollary \ref{equival2}}]
   Using that $S_0(B_C) \le z_{gen}(B_C) \le z_{opt}(B_C)$, we have
   $1 \le \frac{\mathbb{E}_C\left[ z_{gen} (B_C)\right]}{\mathbb{E}_C\left[ S_0 (B_C)\right]}
   \le \frac{\mathbb{E}_C\left[ z_{opt} (B_C)\right]}{\mathbb{E}_C\left[ S_0 (B_C)\right]}
   \le f(C)$. Finally we know that $\mathbb{E}_C\left[ S_0 (B_C)\right]=\alpha_h C$.
 \end{proof}

 \begin{proof}[\textbf{Proof of Proposition \ref{pro1}}]
  We can write $\mathbb{E}[e_1]$ as follows:
 \begin{eqnarray}
 \mathbb{E}[e_1] = (\mathbb{E}[\mathbb{E}[{z}_{p_{ASA}}] - \bar{z}_{p_{ASA}}] ) - (\mathbb{E}[\mathbb{E}[z_{p^*}] - \bar{z}_{p^*}])  +  \mathbb{E}[{\bar{z}}_{p_{ASA}}-\bar{z}_{p^*}]
 \le  \mathbb{E}[|\mathbb{E}[{z}_{p_{ASA}}] - \bar{z}_{p_{ASA}}|] +\mathbb{E}[|\mathbb{E}[z_{p^*}] - \bar{z}_{p^*}|].
 \label{proof1}
 \end{eqnarray}

 The above inequality holds because $ASA^*$ chose $p_{ASA}$, and thus we have $\bar{z}_{p_{ASA}} < \bar{z}_{p^*}$. Moreover, we can compute $\mathbb{E}[|\mathbb{E}[{z}_{p_{ASA}}] - \bar{z}_{p_{ASA}}|]$ and $\mathbb{E}[|\mathbb{E}[z_{p^*}] - \bar{z}_{p^*}|]$ using their $CDF$. We denote $|\mathbb{E}[{z}_{p_{ASA}}] - \bar{z}_{p_{ASA}}|$ by $\Delta$. Note that $0\le\Delta \le r_{max}$ and inequality \eqref{hoeffding} gives a bound on its $CDF$, when estimating $\bar{z}_{p_{ASA}}$ with $S$ samples. For $\mathbb{E}[\Delta]$, we have:

 \begin{eqnarray}
 \mathbb{E}[\Delta]  &=& \int_0^{r_{max}}1 - F_{\Delta}(x) dx \\
 \label{first}
 &\le& \int_0^{r_{max}} 2\exp(\dfrac{-2S {{x^2}}}{{r^2}_{max}}) dx  \\
 &=& 2\int_0^{r_{max}} (\frac{\epsilon}{2}) ^ {\frac{{x^2}}{\delta^2}} dx
 \label{second}
 \le 2\int_0^{\infty} (\frac{\epsilon}{2}) ^ {\frac{{x^2}}{\delta^2}} dx  = \delta \sqrt{\dfrac{\pi}{-\ln({\dfrac{\epsilon}{2}})}}.
 \label{exp_XL}
 \end{eqnarray}

 We can compute $\mathbb{E}[|\mathbb{E}[z_{p^*}] - \bar{z}_{p^*}|]$ in a similar way.

 The first equality in \eqref{second} is obtained by substituting $S$ with the expression in \eqref{numSample}.
 Using the bound on $\mathbb{E}[\Delta]$ and inequality \eqref{proof1}, we can compute $\mathbb{E}[e_1]$ as follows:

 \begin{eqnarray}
 \mathbb{E}[e_1] \le 2 \delta \sqrt{\dfrac{\pi}{-\ln({\dfrac{\epsilon}{2}})}}.
 \label{error1}
 \end{eqnarray}

 \end{proof}

 \begin{proof}[\textbf{Proof of Proposition \ref{pro2}}]
 For an arbitrary initial bay $B$,
 let $\hat{p}$ be the path with the minimum estimated upper bound. Notice that pruning $\tilde{p}$ may be a mistake if the true mean of $L_{\tilde{p}}$ is less than the true mean of $U_{\hat{p}}$, i.e., $\mathbb{E}[{L}_{\tilde{p}}] < \mathbb{E}[U_{\hat{p}}]$; in such a case, $e_2 \le \mathbb{E}[U_{\hat{p}}]-\mathbb{E}[z_{\tilde{p}}]
 \le \mathbb{E}[U_{\hat{p}}]-\mathbb{E}[{L}_{\tilde{p}}]$. Otherwise $e_2$ is zero; thus we have $e_2 \le (\mathbb{E}[U_{\hat{p}}]-\mathbb{E}[{L}_{\tilde{p}}])^+$.

 Let us denote $\overline{L}_{\tilde{p}} - \mathbb{E}[{L}_{\tilde{p}}]$ and $\mathbb{E}[U_{\hat{p}}] - \overline{U}_{\hat{p}}$ by $x_L$ and $x_U$, respectively. Also, let $d$ be $\overline{L}_{\tilde{p}}-\overline{U}_{\hat{p}}$. The true and estimated values of the upper bound and lower bound, and the loss are illustrated in Figure \ref{ublb}; $e_2$ is shown by the thick line segment. For any bay with an arbitrary initial configuration, we can bound $e_2$ as follows:
 \begin{eqnarray}
 e_2 \le \big((\mathbb{E}[U_{\hat{p}}] - \overline{U}_{\hat{p}}) + (\overline{U}_{\hat{p}} - \overline{L}_{\tilde{p}}) + (\overline{L}_{\tilde{p}} - \mathbb{E}[{L}_{\tilde{p}}])\big)^+ = (x_U - d + x_L)^+.
 \label{e_2}
 \end{eqnarray}


 The expected loss, $\mathbb{E}[e_2]$, can be bounded as shown in \eqref{exp_loss2}.

 \begin{eqnarray}
 \label{exp_loss2_line1}
 \mathbb{E}[e_2] = \mathbb{E}[(x_L + x_U - d)^+] &\le& \mathbb{E}[(x_L + x_U)^+] \le \mathbb{E}[|x_L+x_U|] \\
 \label{exp_loss2_line2}
 &\le& \mathbb{E}[|x_L|+|x_U|] = \mathbb{E}[|x_L|] + \mathbb{E}[|x_U|] \\
 \label{exp_loss2_line3}
 &\le& 2\delta \sqrt{\dfrac{\pi}{-\ln({\dfrac{\epsilon}{2}})}}.
 \label{exp_loss2}
 \end{eqnarray}

 The first inequality in \ref{exp_loss2_line1} holds because $d$ is always positive. Inequality \ref{exp_loss2_line2} results from the triangular inequality, and the last inequality is obtained by replacing $\mathbb{E}[|x_L|]$ and $\mathbb{E}[|x_U|]$ by the expression in \eqref{exp_XL}.

 \end{proof}

 \begin{proof}[\textbf{Proof of Proposition \ref{pro3}}]

 For an arbitrary initial bay $B$, let $\hat{p}_{max}$ be the path that maximizes $\mathbb{E}[U_{\hat{p}}]$ among $\hat{p}_1,\dots,\hat{p}_m$.
 We bound $e_3$ (the expected loss due to prununig) as follows:
 \begin{eqnarray}
 \label{e3_line1}
 e_3 &\le& P(\textrm{mistake at } t_1 \cup \dots \cup \textrm{mistake at  } t_m )
 (\mathbb{E}[z_{\hat{p}_{max}}]-\mathbb{E}[z_{p^*}]) \\
 \label{e3_line2}
 &\le& m \textrm{ } P(\textrm{mistake at  } t_1) (\mathbb{E}[U_{\hat{p}_{max}}]-0) \\
 \label{e3_line3}
 &=& m \textrm{ } P(\textrm{mistake at  } t_1) (\mathbb{E}[U_{\hat{p}_{max}}] - \overline{U}_{\hat{p}_{max}} + \overline{U}_{\hat{p}_{max}}) \label{e3_line4}\\
 &\le& m \textrm{ } P(\textrm{mistake at  } t_1) (|x_U| + \overline{U}_{\hat{p}}^{max}).
 \label{e3}
 \end{eqnarray}

 Notice that inequality \ref{e3_line2} is obtained by using the union bound. Moreover, we replace the probability of mistakes at each step by the maximum probability. Without loss of generality we assume that maximum probability is at $t_1$.
 Also note that we do not know which path is $\hat{p}_{max}$ because we have not observed $\mathbb{E}[U_{\hat{p}_1}] , \dots, \mathbb{E}[U_{\hat{p}_m}]$. Nevertheless, regardless of the path,
 we can replace $\mathbb{E}[U_{\hat{p}_{max}}] - \overline{U}_{\hat{p}_{max}}$ with $|x_U|$ (by definition) to obtain \eqref{e3}. Also, we can bound $\overline{U}_{\hat{p}_{max}}$ by $\overline{U}_{\hat{p}}^{max}$.


 Now we compute $P(\textrm{mistake at } t_1)$. Let us denote $|x_L|$ and $|x_U|$ by $x$ and $y$, respectively. Also, let $w=x+y$ and $f_W(w)$ be the $PDF$ of $w$. Notice that the probability of making a mistake depends on the value of $d$; the maximum probability corresponds to the stage with the smallest value of $d$ and can be computed as follows:
 \begin{eqnarray}
 P(\textrm{mistake at  } t_1) = P(x + y - d_{min} > 0) = \int_{d_{min}}^{\infty} f_{W}(w) dw
 = \int_{d_{min}}^{\infty} \int_0^{w} f_{X}(w-y)f_{Y}(y) \textrm{ } dy \textrm{ }dw.
 \end{eqnarray}

 $f_X(x)$ and $f_Y(y)$ can be obtained from \eqref{hoeffding}, and the above integral can be bounded as follows:

 \begin{eqnarray}
 \int_{d_{min}}^{\infty}\int_0^{w} f_{X}(w-y)f_{Y}(y) &=& \int_{d_{min}}^{\infty}\int_0^{w}
 4 \big(\dfrac{-\ln(\frac{\epsilon}{2})}{\delta^2}\big)^2 y (w - y)
 (\dfrac{\epsilon}{2})^\frac{(y^2 + (w - y)^2)}{\delta^2}\nonumber\\
 &=& (\dfrac{\epsilon}{2})^{\frac{d_{min}^2}{\delta^2}}
 + \dfrac{d_{min}}{\delta} \sqrt{\dfrac{-ln(\frac{\epsilon}{2})\pi}{2}}\textrm{ }
 (\dfrac{\epsilon}{2})^{\frac{d_{min}^2}{2\delta^2}}
 \textrm{Erf}(\frac{d_{min}\sqrt{-ln(\frac{\epsilon}{2})}}{\sqrt{2}\delta}) \nonumber\\
 &\le&
 (\dfrac{\epsilon}{2})^{\frac{d_{min}^2}{\delta^2}}
 + \dfrac{d_{min}}{\delta} \sqrt{\dfrac{-ln(\frac{\epsilon}{2})\pi}{2}}\textrm{ }
 (\dfrac{\epsilon}{2})^{\frac{d_{min}^2}{2\delta^2}}.
 \label{e3_3}
 \end{eqnarray}

 %

 From \ref{e3}, \ref{e3_3}, and expression \ref{exp_XL} for $\mathbb{E}[|x_U|]$, it follows that:

 \begin{eqnarray}
 \mathbb{E}[e_3] \le m \textrm{ }\bigg(
 \textrm{ } (\dfrac{\epsilon}{2})^{\frac{d_{min}^2}{\delta^2}} +
 \dfrac{d_{min}}{\delta} \sqrt{\dfrac{-ln(\dfrac{\epsilon}{2})\pi}{2}}\textrm{ } (\dfrac{\epsilon}{2})^{\frac{d_{min}^2}{2\delta^2}}
 \bigg)
 \bigg(
 \delta \sqrt{\dfrac{\pi}{-\ln({\dfrac{\epsilon}{2}})}} + \overline{U}_{\hat{p}}^{max}
 \bigg),
 \\ \nonumber
 \textrm{where} \quad
 \overline{U}_{\hat{p}}^{max} = \max\{\overline{U}_{\hat{p}_1},\dots,\overline{U}_{\hat{p}_m}\},
 \textrm{  and  }
 d_{min} = \min\{d_1,\dots,d_m\}.
 \end{eqnarray}
 \end{proof}
\end{document}